\documentclass[a4paper,11pt]{article}
\pdfoutput=1
\usepackage{jheppub}

\usepackage[utf8]{inputenc}

\usepackage{amssymb}
\usepackage{amsmath}
\usepackage{braket}
\usepackage{xcolor}
\usepackage{graphicx}
\usepackage{slashed}

\allowdisplaybreaks

\newcommand{\e}{\epsilon}

\newcommand{\zero}{{(0)}}
\newcommand{\one}{{(1)}}
\newcommand{\two}{{(2)}}

\newcommand{\als}{\alpha_s}
\newcommand{\alsmu}{\alpha_s(\mu)}
\newcommand{\alspi}{\frac{\alpha_s}{4\pi}}
\newcommand{\alsmupi}{\frac{\alpha_s(\mu)}{4\pi}}

\newcommand{\Ord}{\mathcal{O}}

\newcommand{\nn}{\nonumber}

\newcommand{\df}{d}

\newcommand{\Gcusp}{\Gamma^{\text{cusp}}}
\newcommand{\Lp}{L_\perp}

\newcommand{\bp}{\vec{b}_T}

\def\cB{\mathcal{B}}
\def\cC{\mathcal{C}}

\def\cI{\mathcal{I}}

\preprint{}

\title{Transverse Parton Distribution and Fragmentation Functions at NNLO: the Gluon Case}

\author{Ming-Xing Luo,}
\emailAdd{mingxingluo@zju.edu.cn}
\author{Tong-Zhi Yang,}
\emailAdd{yangtz@zju.edu.cn}
\author{Hua Xing Zhu,}
\emailAdd{zhuhx@zju.edu.cn}
\author{and Yu Jiao Zhu}
\emailAdd{zhuyujiao@zju.edu.cn}

\affiliation{Zhejiang Institute of Modern Physics, Department of Physics, Zhejiang University, Hangzhou, 310027, China}

\abstract{We calculate in this paper the perturbative gluon transverse momentum dependent parton distribution functions~(TMDPDFs) and fragmentation functions~(TMDFFs) using the exponential regulator for rapidity divergences. We obtain results for both unpolarized and linearly polarized distributions through next-to-next-to leading order in strong coupling constant, and through ${\cal O}(\epsilon^2)$ in dimensional regulator~(finding discrepancy for the linearly polarized gluon TMDPDFs with a previous result in the literature). We find a nontrivial momentum conservation sum rule for the linearly polarized component for both TMDPDFs and TMDFFs in the ${\cal N}=1$ super-Yang-Mills theory. The TMDFFs are used to calculate the two-loop gluon jet function for the energy-energy correlator in Higgs gluonic decay in the back-to-back limit.}

\keywords{SCET, TMDs, beam function, fragmentation function, QCD corrections}

\begin{document}

\maketitle

\clearpage

\section{Introduction}
\label{sec:introduction}

Transverse Momentum Dependent Parton Distribution Functions~(TMDPDFs) and Fragmentation Functions~(TMDFFs) probe the transverse structure of hadrons. Analytical calculations, phenomenological applications, and experimental determinations of the TMD distributions play important role in understanding the structure of hadrons~\cite{Collins:2011zzd,Angeles-Martinez:2015sea}. 

TMDPDFs and TMDFFs have important applications in collider processes, such as Drell-Yan~\cite{Dokshitzer:1978yd,Parisi:1979se,Collins:1984kg,Arnold:1990yk,Ladinsky:1993zn,Bozzi:2010xn,Becher:2011xn,Bizon:2019zgf,Bertone:2019nxa} and Higgs production~\cite{Berger:2002ut,Bozzi:2005wk,Gao:2005iu,Echevarria:2015uaa,Neill:2015roa,Bizon:2017rah,Chen:2018pzu,Bizon:2018foh}, top quark pair production~\cite{Zhu:2012ts,Li:2013mia,Catani:2014qha,Catani:2018mei}, hadronic $J/\psi$ production, semi-inclusive deep-inelastic scattering~\cite{Ji:2004wu,Ji:2004xq,Su:2014wpa,Kang:2015msa,Liu:2018trl}, hadron or jet production in electron-positron annihilation~\cite{Collins:1981uk,Collins:1981va,Neill:2016vbi,Gutierrez-Reyes:2018qez,Gutierrez-Reyes:2019vbx,Gutierrez-Reyes:2019msa}, and energy correlators in both $e^+e^-$ and hadron colliders~\cite{Collins:1981uk,Moult:2018jzp,Gao:2019ojf}. The TMDPDFs and TMDFFs are intrinsically non-perturbative objects. However, at low transverse momentum but not yet into the non-perturbative region, the TMDPDFs and TMDFFs admit light-cone operator product expansion onto the usual collinear PDFs and FFs, with perturbative calculable matching coefficients. Knowledge for these matching coefficients at higher orders are essential for achieving precision predictions.

In this paper, we present the results for the perturbative matching coefficients at Next-to-Next-to Leading Orders~(NNLO) for gluon TMDPDFs and TMDFFs. They are relevant to the production and decay of the Higgs boson, top quark pair, and $J/\psi$ at low transverse momentum at NNLO or Nex-to-Next-to-Next-to Leading Logarithms~(N$^3$LL) order. We also provide the bare results at NNLO through $\Ord(\e^2)$, which do not contribute to the renormalized TMD coefficients at this order, but are relevant for future N$^3$LO calculation. 
It is well-known that, direct calculation of the TMD matching coefficients requires some form of regularization in addition to the usual dimensional regularization~\cite{Collins:1984kg,Ji:2004wu,Collins:2011zzd,Becher:2010tm,Becher:2011dz,Chiu:2012ir,Chiu:2009yx,Echevarria:2015byo,Li:2016axz,Ebert:2018gsn,Liu:2019iml}. In this paper, we adopt the exponential regularization scheme for the rapidity divergences~\cite{Li:2016axz}.

For the gluon TMD coefficients, the non-trivial spin structure leads to two independent tensor structures in transverse impact parameter space. They are known as the unpolarized coefficients and linearly polarized coefficients~\cite{Mulders:2000sh}. The linearly polarized coefficients arise from helicity-flip contribution in gluon splitting, and are suppressed by one power of $\alpha_s$ compared with the unpolarized ones. Their contributions to physical cross section have been discussed for diphoton production~\cite{Nadolsky:2007ba}, Higgs production~\cite{Catani:2010pd,Boer:2011kf,Boer:2013fca,Echevarria:2015uaa}, quarkonium production~\cite{Dunnen:2014eta,Mukherjee:2016cjw,Lansberg:2017dzg}, $\gamma^*$ plus jet production~\cite{Dominguez:2011br,Boer:2017xpy}, heavy quark pair~\cite{Pisano:2013cya} and dijet production~\cite{Boer:2016fqd}. In this paper we present results through $\Ord(\alpha_s^2)$~(two loops) for both polarizations. For the linear polarization, this is formally at the NLO accuracy, since its LO contribution already starts at $O(\alpha_s)$. We note that with the two-loop linearly polarized contributions presented in this paper, the linearly polarized ingredients needed for N$^3$LO calculation~\cite{Bizon:2018foh,Billis:2019vxg} for Higgs production using the $q_T$ subtraction formalism~\cite{Catani:2007vq} are completed. The reason is that the pure three-loop linearly polarized component will not contribute to the cross section at $\Ord(\alpha_s^3)$, since the tree-level linear polarized contribution vanishes, and its interference with the unpolarized contribution vanishes. 

As a by-product of this calculation, we find an interesting momentum conservation sum rule for the linearly polarized TMD coefficients in the ${\cal N}=1$ supersymmetric limit. The sum rule imposes non-trivial constraint to the gluon-to-gluon and quark-to-gluon TMD coefficients, and are found to be satisfied by our two-loop results. 

The unpolarized gluon TMDPDFs have been computed before using different regulators for the rapidity divergences. Our results can be compared with them when combining with the rapidity-regulator dependent TMD soft function properly~\cite{Li:2016ctv}. We have done this exercise and found complete agreement with Refs.~\cite{Gehrmann:2014yya,Echevarria:2015usa}. For unpolarized gluon TMDFFs, results have also been given through NNLO~\cite{Echevarria:2016scs}. Our results agree with them for most of the terms, except for a term of the form $C_A^2 \pi^4 \delta(1-z)$ in the gluon-to-gluon TMD coefficient. We have also found similar discrepancy in the quark-to-quark coefficient as reported in Ref.~\cite{Luo:2019hmp}. Very recently, the two-loop results for the linearly polarized gluon TMDPDFs have been given in Ref.~\cite{Gutierrez-Reyes:2019rug}. We have compared our results with Ref.~\cite{Gutierrez-Reyes:2019rug} by constructing rapidity-regulator independent TMDPDFs, and found discrepancy with Ref.~\cite{Gutierrez-Reyes:2019rug}. We have performed several checks to our calculations, which will be explained below. 

This paper is organized as follows. In Sec.~\ref{sec:gluon-tmdpdf}, we give the bare and renormalized results for the gluon TMDPDFs through NNLO. In Sec.~\ref{sec:gluon-tmdff}, we give the bare and renormalized results for the gluon TMDFFs through NNLO. In Sec.~\ref{sec:gluon-jet-funcion} we use the gluon TMDFFs to calculate the gluon jet function, which is relevant for the Energy-Energy Correlations for Higgs gluonic decay in the back-to-back limit. We conclude in Sec.~\ref{sec:conclusion}. We collect the relevant perturbative ingredients in the appedix~\ref{sec:appendix}.  

\section{Gluon TMDPDFs}
\label{sec:gluon-tmdpdf}
The bare gluon TMDPDF can be defined in terms of SCET~\cite{Bauer:2000ew,Bauer:2000yr,Bauer:2001yt,Bauer:2002nz,Beneke:2002ph} collinear gauge fields
\begin{align}
  \label{eq:gPDFdef}
  {\cal B}_{g/N}^{\rm bare, \mu\nu}(x,b_\perp) = - x P_+ \int \frac{d b_-}{4 \pi} e^{- i x b_- P_+/2} \langle N (P) | {\cal A}_{n \perp}^{a,\mu} (0, b_-, b_\perp) {\cal A}_{n \perp}^{a,\nu}(0) | N(P) \rangle \,,
\end{align}
where $A_{n\perp}^{a,\mu}$ is the gauge invariant collinear gluon field with color index $a$ and Lorentz index $\mu$. For sufficiently small $b_\perp$, the gluon TMDPDFs admit operator production expansion onto the usual collinear PDFs,
\begin{align}
  \label{eq:PDFOPE}
{\cal B}_{g/N}^{\rm bare, \mu\nu}(x,b_\perp) = \sum_i \int_x^1 \frac{d \xi}{\xi} {\cal I}_{gi}^{{\rm bare}, \mu\nu} (\xi, b_\perp) \phi_{i/N}^{\rm bare}(x/\xi) + \Ord(b_T^2 \Lambda_{\rm QCD}^2 ) \,,
\end{align}
where the summation is over all parton flavors $i$. The coefficient functions can be decomposed into two independent Lorentz structures,
\begin{align}
  \label{eq:PDFdecomposition}
  {\cal I}_{gi}^{{\rm bare}, \mu\nu} (\xi, b_\perp) = \frac{g_\perp^{\mu\nu}}{d-2} 
{\cal I}_{gi}^{{\rm bare}}(\xi, b_T) + \left(\frac{g_\perp^{\mu\nu}}{d-2} + \frac{b_\perp^\mu b_\perp^\nu}{b_T^2} \right) 
{\cal I'}_{gi}^{{\rm bare}}(\xi, b_T) \,,
\end{align}
where we have defined two scalar form factor ${\cal I}_{gi}^{\rm bare}$ and ${\cal I}^{'\rm bare}_{gi}$, which can be projected out using
\begin{align}
\label{eq:PDFProjection}
\mathcal{I}^{{\rm bare}}_{gi}(\xi, b_T) &= g_\perp^{ \mu \nu} {\cal I}_{gi}^{{\rm bare}, \mu\nu} (\xi, b_\perp)\,, \nonumber \\
\mathcal{I'}^{{\rm bare}}_{gi}(\xi, b_T) &=  \frac{1}{d-3} \left[  g_\perp^{\mu \nu} + \left(   d-2\right) \frac{b_\perp^\mu b_\perp^\nu}{b_T^2} \right] {\cal I}_{gi}^{{\rm bare}, \mu\nu} (\xi, b_\perp) \,, 
\end{align}
with $ b_T^2 = -b_\perp^2$ and $b_T = \sqrt{b_T^2} \,.$ 

The matching coefficients $\cI^{\rm bare, \mu \nu}_{gi}(\xi, b_\perp)$ in Eq.~\eqref{eq:PDFOPE} do not depend on the actual hadron $N$. In actual calculation, one can replace the hadron $N$ with a partonic state $j$. Furthermore, the usual bare partonic collinear PDFs are just $\phi^{\rm bare}_{i/j}(x) = \delta_{ij} \delta(1-x)$, so one has 
\begin{align}
\cI^{\rm bare, \mu \nu}_{gi}(x, b_\perp) = \cB^{\rm bare , \mu \nu}_{gi}(x, b_\perp) \,. 
\end{align}
The TMDPDFs, as well as their matching coefficients, contain both UV and rapidity divergences. We adopt dimensional regularization for the UV, and exponentional regularization~\cite{Li:2016axz} for the rapidity divergences. In the following subsection, we present the bare results for the coefficient functions through two loops in QCD.

\subsection{The bare results}
\label{sec:bare-results}

The relevant diagrams for gluon TMDPDF through two loops are generated with the code \textsc{Qgraf}~\cite{Nogueira:1991ex}. We use an in-house \textsc{Mathematica} code to substitute in the SCET Feynman rules, and use \textsc{Form}~\cite{Vermaseren:2000nd} to carry out necessary color and algebra manipulation. We employ reverse unitarity~\cite{Anastasiou:2002yz} to convert phase space integral into loop integral for the purpose of integral reduction. We use \textsc{Fire5}~\cite{Smirnov:2014hma} and \textsc{LiteRed}~\cite{Lee:2012cn} to reduce the integrand into the so-called master integrals by Integration-By-Parts identities~\cite{Chetyrkin:1981qh}. The resulting master integrals have been solved in Ref.~\cite{Luo:2019hmp} using differential equation method~\cite{Gehrmann:1999as,Henn:2013pwa}. For the details we refer to Ref.~\cite{Luo:2019hmp}.

The bare coefficient functions can be expanded in terms of QCD bare coupling as
\begin{align}
  \label{eq:bare_expansion}
  {\cal I}_{gi}^{{\rm bare}, \mu\nu} (\xi, b_\perp, \mu, \nu, \e, \alpha_0)
= \sum_{k=0}^{\infty} \left(\frac{\alpha_0 S_\epsilon}{4 \pi} \right)^k {\cal I}_{gi}^{(k) {\rm bare}, \mu\nu}(\xi, b_\perp, \mu, \nu, \e) \,,
\end{align}
and similarly for the form factor ${\cal I}_{gi}^{\rm bare}$ and ${\cal I'}_{gi}^{\rm bare}$, where $S_\epsilon = \left( 4 \pi e^{- \gamma_E } \mu_0^2/\mu^2 \right)^\epsilon $. The one-loop bare coefficient functions up to $\mathcal{O}(\epsilon^2)$ are
\begin{align}
    \mathcal{I}^{'(1) \text{bare}}_{gq} &= 2 C_F r_b(x) \biggl(  2 + 2 L_\perp  \epsilon + (  L_\perp^2 + \zeta_2)\epsilon^2  \biggl)   \,,  \nonumber \\
    \mathcal{I}^{'(1) \text{bare}}_{gg} &= 2 C_A r_b(x) \biggl(  2 + 2 L_\perp  \epsilon + (  L_\perp^2 + \zeta_2)\epsilon^2  \biggl) \,,  \nonumber \\
  \mathcal{I}^{(1) \text{bare}}_{gq} &= -\frac{2 C_F p_{gq}(x)}{\epsilon }+C_F\left(2x-2p_{gq}(x) L_\perp \right)+\epsilon  C_F \left(-p_{gq}(x) L_\perp^2-\zeta_2 p_{gq}(x)+2 x L_\perp \right) \nonumber \\
   &+\epsilon ^2 C_F
   \left(-\zeta _2p_{gq}(x) L_\perp-\frac{1}{3 }p_{gq}(x) L_\perp^3-\frac{2 \zeta _3
  p_{gq}(x)}{3}+x L_\perp^2+x \zeta _2\right)  \,, \nn \\
\mathcal{I}^{(1) \text{bare}}_{gg}&= \frac{C_A}{\epsilon^2} \biggl(4 \delta(1-x)\biggl) + \frac{C_A}{\epsilon}  \biggl(-2 \delta (1-x) L_{\nu  Q}-4 p_{gg}(x)\biggl) \nonumber \\
&+   C_A \bigg\{ \delta (1-x) \left(-2 L_\perp L_{\nu  Q}-2 L_\perp^2-2 \zeta _2\right)-4 p_{gg}(x)
   L_\perp     \bigg\}   \nonumber \\
   &+ \epsilon  C_A \bigg\{\delta (1-x) \left(-L_\perp^2 L_{\nu  Q}-4 \zeta _2 L_\perp-\frac{4
   L_\perp^3}{3}-\zeta _2 L_{\nu  Q}-\frac{8 \zeta _3}{3}\right) \nonumber \\
   &+p_{gg}(x) \left(-2
   L_\perp^2-2 \zeta _2\right)\bigg\} + \epsilon ^2 C_A \bigg\{   \delta (1-x) \biggl(-\frac{1}{3} L_\perp^3 L_{\nu  Q}+L_\perp \left(-\zeta _2 L_{\nu 
   Q}-4 \zeta _3\right) \nonumber \\
   &-3 \zeta _2 L_\perp^2-\frac{L_\perp^4}{2}-\frac{2}{3} \zeta _3
   L_{\nu  Q}-\frac{27 \zeta _4}{4}\biggl)+     p_{gg}(x) \left(-2 \zeta _2 L_\perp-\frac{2 L_\perp^3}{3}-\frac{4 \zeta _3}{3}\right)   \bigg\}  \,,
 \end{align}
where we have defined
\begin{align}
\label{eq:rb}
r_b(x) = \frac{1-x}{x}, \quad  L_{\nu Q} =2 L_\nu + L_Q \,,  
\end{align}
and 
\begin{align}
\label{eq:Ldefinition}
 L_\perp = \ln \frac{b_T^2 \mu^2}{b_0^2} , \quad  L_Q  = 2 \ln \frac{x \, P_+}{\nu}, \quad L_\nu = \ln \frac{\nu^2}{\mu^2} \,,
\end{align}
with $b_0 =2 \, e^{- \gamma_E}$. Our notations are the same as in Ref.~\cite{Luo:2019hmp}.

The two-loop bare results up to $\Ord(\epsilon^0)$ are
\begin{align}
   \mathcal{I}^{'(2) \text{bare}}_{gq} &= \frac{16 C_A C_F}{\epsilon^2} r_b(x) + \frac{1}{\epsilon} \bigg\{  C_F^2 \bigg[16 H_1 r_b(x)-8 H_0-4 (1-x)\bigg]-\frac{32}{3} C_F N_f T_F r_b(x) \nonumber \\
   & + C_A C_F \bigg[\frac{1}{3} r_b(x) \biggl(48 H_0+48 L_\perp-24 L_{\nu  Q}+4 (3x+43)\biggl)+32 H_0\bigg]    \bigg\} \nonumber \\
   & + \epsilon^0 \bigg\{  C_A C_F \bigg[ \frac{1}{3} r_b(x) \biggl(48 H_{1,1}+L_\perp \left(96 H_0-48 L_{\nu  Q}+8 (3
   x+43)\right)-40 H_1 \nonumber \\
   &+48 H_2+\frac{8}{3} (9 x-11)-48 \zeta _2\biggl) + 16 H_{0,0}+64  L_\perp H_0-\frac{8  (5 x+2)}{x} H_0    \bigg] \nonumber \\
   & + C_F^2 \bigg[ \frac{1}{3} r_b(x) \left(-48 H_{1,1}+96  L_\perp H_1 +24 H_1\right)-8 H_{0,0}+\left(-16
   H_0-8 (1-x)\right) L_\perp  \nonumber \\
   & +20 H_0-8 (1-x)  \bigg] +\frac{1}{3} C_F N_f T_F r_b(x) \left(32 H_1-64 L_\perp-\frac{64}{3}\right)    \bigg\} +\Ord\left(\epsilon ^1\right) \,, \nonumber \\
  \mathcal{I}^{'(2) \text{bare}}_{gg} &= \frac{16 C_A^2}{\epsilon^2} r_b(x)  + \frac{1}{\epsilon} \bigg\{  C_F N_f T_F \left(\frac{16}{3} \left(x^2-2 x-2\right) r_b(x)-16
   H_0\right)-\frac{16}{3} C_A N_f T_F r_b(x)  \nonumber  \\
   & + C_A^2 \left(\frac{1}{3} r_b(x) \left(48 H_0+48 H_1+48 L_\perp-24 L_{\nu  Q}-4 \left(2
   x^2-4 x-45\right)\right)+32 H_0\right) \bigg\}  \nonumber \\
   &+ \epsilon^0 \bigg\{ C_A^2 \bigg[ \frac{1}{3} r_b(x) \biggl(L_\perp \left(96 H_0+96 H_1-48 L_{\nu  Q}-8 \left(2 x^2-4
   x-45\right)\right)+48 H_2 \nonumber \\ 
   &-48 \zeta_2\biggl) + 16 H_{0,0}+64  L_\perp H_0-\frac{4 (19 x+12)}{3 x} H_0 +\frac{4 \left(11 x^3-39 x^2+62
   x-37\right)}{9 x}  \bigg] \nonumber \\
   &+ C_F N_f T_F \bigg[-16 H_{0,0}-32  L_\perp H_0 +\frac{32}{3} \left(x^2-2 x-2\right) L_\perp
   r_b(x) +\frac{16 (1-x)^3}{x}\bigg] \nonumber \\
   &+ C_A N_f T_F \bigg[ -\frac{16 H_0}{3}-\frac{32 L_\perp r_b(x)}{3}+\frac{8 \left(x^3+3 x^2+16
   x-17\right)}{9 x}    \bigg]    \bigg\} +O\left(\epsilon ^1\right) \,, \nonumber \\
    \mathcal{I}^{(2) \text{bare}}_{gq} & =  \frac{-8 C_A C_F}{\epsilon^3} p_{gq}(x)  + \frac{1}{\epsilon^2} \bigg\{C_F^2 \bigg[-4  p_{gq}(x) H_1 +2 (2-x) H_0 -x+4 \bigg]   \nonumber \\
    & + \frac{8}{3} C_F N_f T_F p_{gq}(x)  + C_A C_F \bigg[ -4 p_{gq}(x) H_1 +4 p_{gq}(x) L_{\nu  Q}-8 p_{gq}(x) L_\perp \nonumber \\
    &-\frac{8  \left(x^2+x+1\right)}{x} H_0 +\frac{2 \left(4 x^3+4 x^2+46 x-53\right)}{3 x}  \bigg]     \bigg\} \nonumber \\
    & + \frac{1}{\epsilon} \bigg\{  C_F N_f T_F \bigg[-\frac{8  p_{gq}(x) }{3} H_1 +\frac{16  p_{gq}(x)
   L_\perp}{3}+\frac{16 \left(x^2-5 x+5\right)}{9 x}\bigg] \nonumber \\
   & + C_F^2 \bigg[ 4 H_{1,1} p_{gq}(x)+2 (2-x) H_{0,0}+L_\perp \left(-8 H_1 p_{gq}(x)+4 H_0 (2-x)+2 (4-x)\right) \nonumber \\
   & -\frac{2 \left(x^2-6 x+6\right)}{x} H_1  + (-3 x-4) H_0 +5
   x+1\bigg] + C_A C_F \bigg[ L_\perp \biggl(-8  p_{gq}(x) H_1 \nonumber \\
   &+8 p_{gq}(x) L_{\nu  Q}-\frac{16
   \left(x^2+x+1\right)}{x} H_0 +\frac{4 \left(4 x^3-2 x^2+46 x-53\right)}{3 x}\biggl) - 4 x L_{\nu Q}  \nonumber \\
   & + 4 p_{gq}(-x)  H_{-1,0} +4 \left(-H_{1,0}-H_{1,1}-H_2\right) p_{gq}(x)-4 (x+2)
   H_{0,0} -8 \zeta _2 \nonumber \\
   &+\frac{2}{3}\left(8 x^2+15 x+36\right)  H_0  +\frac{2 \left(17 x^2-22
   x+22\right)}{3 x}  H_1 \nonumber \\
   &  -\frac{2 \left(44 x^3-29 x^2+19 x+9\right)}{9 x}    \bigg]     \bigg\} + \epsilon^0 \bigg\{  C_A C_F \bigg[ \frac{16}{3} L_\perp^3 p_{gq}(x) + L_\perp^2 \biggl( -8 H_1 p_{gq}(x) \nonumber \\
   & +8 p_{gq}(x) L_{\nu  Q}-\frac{16 H_0
   \left(x^2+x+1\right)}{x}+\frac{4 \left(4 x^3-8 x^2+46 x-53\right)}{3 x}   \biggl)  \nonumber \\
   &+ L_\perp \biggl( 8  p_{gq}(-x) H_{-1,0}+4 \left(-2 H_{1,0}-2 H_{1,1}-2 H_2\right) p_{gq}(x)-8
   (x+2) H_{0,0} \nonumber \\
   &+\frac{4}{3}\left(8 x^2+15 x+36\right) H_0+\frac{4  \left(17 x^2-22
   x+22\right)}{3 x} H_1 -8 x L_{\nu  Q} \nonumber \\
   &+\frac{8 \left(x^2-4 x+2\right) \zeta _2}{x}-\frac{4
   \left(44 x^3-29 x^2+19 x+9\right)}{9 x}  \biggl) + 4 \zeta_2 p_{gq}(x) L_{\nu Q}  \nonumber \\
   &-4 p_{gq}(-x) \left(-2 H_{-2,0}+2 H_{-1,-1,0}-H_{-1,0,0}+H_{-1} \zeta _2\right)  \nonumber \\
   &+4p_{gq}(x) \left(H_{1,2}+H_{2,1}-H_{1,0,0}+H_{1,1,0}+H_{1,1,1}-H_1 \zeta
   _2\right) \nonumber \\
   &+\frac{2}{3} \left(8 x^2+9 x+36\right) H_{0,0}-\frac{2 \left(5 x^2-22
   x+22\right)}{3 x}  H_{1,1} +\frac{8 \left(x^2+2\right) }{x}H_{2,0} \nonumber \\
   &-\frac{4 \left(4 x^3-9x^2+24 x-22\right) }{3 x} H_{1,0}+4 x H_{-1,0}-4 (x+2) H_{0,0,0}\nonumber \\
   &-\frac{8  \left(x^2+x+1\right) \zeta _2}{x} H_0 -\frac{2}{9} H_0 \left(88 x^2-6 x+249\right)+\frac{2
    \left(43 x^2-152 x+152\right)}{9 x}H_1 \nonumber \\
    & + 4 x H_2+\frac{8 \left(10 x^2-23 x+20\right)
   \zeta _3}{3 x}-\frac{2 \left(4 x^3-4 x^2+2 x+9\right) \zeta _2}{3 x} \nonumber \\
   & +\frac{4 \left(152 x^3-268 x^2+791 x-790\right)}{27 x}   \bigg] + C_F^2 \bigg[ L_\perp^2 \biggl(-8  p_{gq}(x) H_1 +4  (2-x)H_0 \nonumber \\
   & +2 (4-x)\biggl) + L_\perp \biggl(8 H_{1,1} p_{gq}(x)+4 (2-x) H_{0,0}-\frac{4  \left(x^2-6
   x+6\right)}{x}H_1 \nonumber \\
   &-2(3 x+4) H_0 +2 (5 x+1)\biggl) + 4 p_{gq}(x) \left(-H_{1,1,1}-H_1 \zeta _2\right)+\frac{2 \left(x^2-6 x+6\right)
  }{x}  H_{1,1} \nonumber \\
  & +2 (2-x) \left(H_{0,0,0}+H_0 \zeta _2\right)+(-3 x-4) H_{0,0}-\frac{2 
   \left(5 x^2-16 x+16\right)}{x} H_1 \nonumber \\
   & +5 (x-3) H_0+(4-x) \zeta _2-x+10    \bigg] + C_F N_f T_F \bigg[ \frac{16}{3} p_{gq}(x) L_\perp^2  \nonumber \\
   &+ L_\perp \left(\frac{32 \left(x^2-5 x+5\right)}{9 x}-\frac{16
   p_{gq}(x)}{3} H_1 \right) + \frac{8}{3} p_{gq}(x) \left( H_{1,1}+ \zeta _2\right) \nonumber \\
   &-\frac{16
   \left(x^2-5 x+5\right)}{9 x} H_1 +\frac{8 \left(13 x^2-56 x+56\right)}{27 x}  \bigg]   \bigg\}+\Ord\left(\epsilon ^1\right) \,,  \nonumber \\  
\mathcal{I}^{(2)\text{bare}}_{gg}&=\frac{1}{\epsilon^4}\bigg\{8 C_A^2 \delta (1-x)\bigg\}+\frac{1}{\epsilon ^3}\bigg\{\delta (1-x)  \left( \frac{11 C_A^2 - 4 C_A N_f T_F }{3}-8 L_{\nu  Q} C_A^2 \right) \nonumber \\ 
  &-16 C_A^2 p_{gg}(x)\bigg\}+ \frac{1}{\epsilon
   ^2}\bigg\{C_A^2 \bigg[16 \left(\frac{\ln (1-x)}{1-x}\right)_++8
   p_{gg}(x) L_{\nu  Q}-16 p_{gg}(x) L_\perp  \nonumber \\
   &-\frac{22 p_{gg}(x)}{3}+\frac{16
   \left(x^3-x^2+2 x-1\right)}{x}  H_1 -\frac{8  \left(x^4-4 x^3+3 x^2+1\right)}{(1-x)
   x} H_0  \nonumber \\ 
   &-\frac{8 (1-x) \left(11 x^2+2 x+11\right)}{3 x}\bigg]+\frac{8}{3} C_A N_f T_F
   p_{gg}(x) \nonumber \\  
  & +\delta (1-x) \bigg[C_A N_f T_F \left(\frac{4 L_{\nu 
   Q}}{3}-\frac{20}{9}\right)+C_A^2 \biggl(-8 L_\perp L_{\nu  Q}-8 L_\perp^2+2 L_{\nu Q}^2 +\frac{67}{9}   \nonumber \\ 
   &-18 \zeta _2-\frac{11 L_{\nu  Q}}{3} \biggl) \bigg] +C_F N_f T_F
   \bigg[8  (x+1)H_0+\frac{4 (1-x) \left(4 x^2+7 x+4\right)}{3 x}\bigg] \bigg\}  \nonumber \\
   &+\frac{1}{\epsilon}\bigg\{C_A^2 \bigg[L_\perp \biggl(32 \left(\frac{\ln (1-x)}{1-x}\right)_++16p_{gg}(x) L_{\nu  Q}+\frac{32  \left(x^3-x^2+2x-1\right)}{x} H_1 \nonumber \\
   &-\frac{16 \left(x^4-4 x^3+3 x^2+1 \right)}{(1-x) x} H_0-\frac{16 (1-x)\left(11 x^2+2 x+11\right)}{3 x} -\frac{44 p_{gg}(x)}{3}\biggl)\nonumber \\
   & +4 p_{gg}(-x) \left(2 H_{-1,0}+\zeta_2\right)-\frac{8\left(x^2-x-1\right)^2 }{(1-x) (x+1)} H_{0,0} +\frac{2}{3}
    \left(44 x^2-11 x+25\right) H_0 \nonumber \\
    &-\frac{1}{9} p_{gg}(x) ( 72 H_{1,0}+72 H_2-36 \zeta _2+134)+\frac{(1-x) \left(134 x^2-109 x+134\right)}{9
   x}\bigg] \nonumber \\
   &+C_F N_f T_F \bigg[8 (x+1) H_{0,0}+\left(16 (x+1) H_0 +\frac{8 (1-x) \left(4
   x^2+7 x+4\right)}{3 x}\right) L_\perp \nonumber \\
   &+4 (x+3)H_0-\frac{8 (1-x) \left(x^2-8 x+1\right)}{3
   x}\bigg]+C_A N_f T_F \bigg[\frac{16 p_{gg}(x) L_\perp}{3}+\frac{40
   p_{gg}(x)}{9}\nonumber \\
   &+\frac{8}{3}  (x+1)H_0 +\frac{4 (1-x) \left(13 x^2+4 x+13\right)}{9
   x}\bigg]+\delta (1-x) \bigg[C_A N_f T_F \biggl(\frac{8}{3} L_\perp L_{\nu  Q} \nonumber \\
   &+\frac{8L_\perp^2}{3}+\frac{20 L_{\nu  Q}}{9}+4 \zeta _2-\frac{112}{27}\biggl)+C_A^2 \biggl(L_\perp
   \left(4 L_{\nu  Q}^2-\frac{22 L_{\nu  Q}}{3}-32 \zeta _2\right)-\frac{16
   L_\perp^3}{3}\nonumber \\
   &-\frac{22 L_\perp^2}{3}+\left(2 \zeta _2-\frac{67}{9}\right) L_{\nu  Q}-11 \zeta_2-\frac{74 \zeta _3}{3}+\frac{404}{27}\biggl)\bigg]\bigg\} \nonumber \\
   &+ \epsilon^0 \bigg\{  C_A^2 \bigg[ \frac{32}{3} p_{gg}(x) L_\perp^3+ L_\perp^2 \biggl(32 \left(\frac{\ln (1-x)}{1-x}\right)_+ +\frac{32  \left(x^3-x^2+2 x-1\right)}{x} H_1 \nonumber \\
   &+16 p_{gg}(x) L_{\nu 
   Q} -\frac{16\left(x^4-4 x^3+3 x^2+1\right)}{(1-x) x} H_0 -\frac{16 (1-x) \left(11 x^2+2
   x+11\right)}{3 x} \nonumber \\
   &-\frac{44 p_{gg}(x) }{3} \biggl) + L_\perp \biggl(-8 p_{gg}(-x) \left(-2 H_{-1,0}-\zeta _2\right)-8 p_{gg}(x) \left(2
   H_{1,0}+2 H_2-3 \zeta _2\right)\nonumber \\
   &-\frac{16 \left(x^2-x-1\right)^2}{(1-x)
   (x+1)}  H_{0,0} -\frac{268 p_{gg}(x) }{9}+\frac{4}{3}  \left(44 x^2-11
   x+25\right)H_0 \nonumber \\
   &+\frac{2 (1-x) \left(134 x^2-109 x+134\right)}{9 x}\biggl) +8 \zeta _2 p_{gg}(x) L_{\nu  Q}    + 16 \zeta _2 \left(\frac{\ln (1-x)}{1-x}\right)_+ \nonumber \\
   &-8 p_{gg}(-x) \left(-2 H_{-2,0}+2
   H_{-1,-1,0}-H_{-1,0,0}+H_{-1} \zeta _2\right)\nonumber \\
   &-8 p_{gg}(x)
   \left(-H_{1,2}-H_{2,1}+H_{1,0,0}-H_{1,1,0}-\frac{25 \zeta _3}{6}\right)+\frac{2}{3}
   \left(44 x^2-11 x+25\right) H_{0,0}\nonumber \\
   &+\frac{8 (1-x) \left(11 x^2-x+11\right) H_{1,0}}{3
   x}- \frac{1}{(1-x)(x+1)}\biggl( 8 \left(x^2-x-1\right)^2 H_{0,0,0}\nonumber \\
   &+\frac{16 \left(x^4-x^2-1\right)
   }{x} H_{2,0} -\frac{4 \left(7 x^5-5 x^4+7 x^3+5 x^2-7 x+5\right) \zeta _3}{x}\biggl)-\frac{22 \zeta _2 p_{gg}(x)}{3} \nonumber \\
   &-\frac{808 p_{gg}(x)}{27}+\frac{1}{9}
    \left(-536 x^2-149 x-701\right)H_0+\frac{16  \left(x^3-x^2+2 x-1\right) \zeta
   _2}{x} H_1 \nonumber \\
   &-\frac{8\left(x^4-4 x^3+3 x^2+1\right) \zeta _2}{(1-x) x}  H_0 -\frac{2 
   x}{3}H_1 +\frac{4 \left(211 x^3-186 x^2+174 x-196\right)}{9 x}\nonumber\\
   &-8 (1-x) \zeta _2\bigg]  + C_A N_f T_F \bigg[\frac{16}{3} p_{gg}(x) L_\perp^2 + L_\perp \biggl(\frac{80 p_{gg}(x)}{9}+\frac{16}{3}  (x+1)H_0 \nonumber \\
   &+\frac{8 (1-x) \left(13 x^2+4
   x+13\right)}{9 x}\biggl) + \frac{8}{3} (x+1) H_{0,0}+\frac{8 \zeta _2 p_{gg}(x)}{3}+\frac{224
   p_{gg}(x)}{27}\nonumber \\
   &+\frac{4}{9} (10 x+13)H_0 +\frac{4  x}{3}H_1-\frac{4 \left(83
   x^3-54 x^2+54 x-65\right)}{27 x}    \bigg] \nonumber \\
   &+ C_F N_f T_F \bigg[ L_\perp^2 \left(16(x+1)  H_0  +\frac{8 (1-x) \left(4 x^2+7 x+4\right)}{3 x}\right) + L_\perp \biggl(16 (x+1) H_{0,0} \nonumber \\
   &+8 (x+3)H_0 -\frac{16 (1-x) \left(x^2-8 x+1\right)}{3 x}\biggl) + 8 (x+1) \left(H_{0,0,0}+H_0 \zeta _2\right) \nonumber \\
   &+4 (x+3) H_{0,0}+24 H_0 (x+1)+\frac{4 (1-x)
   \left(4 x^2+7 x+4\right) \zeta _2}{3 x} \nonumber \\
   &-\frac{8 (1-x) \left(x^2-23 x+1\right)}{3 x} \bigg] + \delta(1-x) \bigg[  C_A^2 \biggl(  \frac{1}{9} L_\perp^3 \left(48 L_{\nu  Q}-88\right) \nonumber \\
   & + \frac{1}{9} L_\perp^2 \left(36 L_{\nu  Q}^2-66 L_{\nu  Q} -180 \zeta _2-134 \right)  +  L_\perp \left( \frac{1}{9}\left(108 \zeta _2-134\right) L_{\nu  Q}-\frac{88}{3} \zeta _2-16 \zeta _3\right)  \nonumber \\
  &+ 2 \zeta _2 L_{\nu  Q}^2 + \left(-\frac{11 \zeta _2}{3}+\frac{50 \zeta _3}{3}-\frac{404}{27}\right) L_{\nu  Q}  -\frac{67 \zeta _2}{3}-\frac{242 \zeta _3}{9}-32 \zeta _4+\frac{2428}{81} \biggl)  \nonumber \\
  &+ C_A N_f T_F \biggl(  \frac{32}{9} L_\perp^3 + \frac{1}{9} L_\perp^2 ( 40 + 24 L_{\nu Q})  + \frac{1}{9} L_\perp ( 40 L_{\nu Q} + 96 \zeta_2) \nonumber \\
  &+ \frac{1}{27} L_{\nu Q}( 112+ 36 \zeta_2) + \frac{20 \zeta _2}{3}+\frac{88 \zeta _3}{9}-\frac{656}{81} \biggl) \bigg]  \bigg\} +\Ord\left(\epsilon ^1\right) \,,
\end{align}
where all end point divergences $ \ln^k (1-x)/(1-x)$ should be taken as plus-distribution, which is defined as 
\begin{align}
\label{eq:Plusdef}
\int_0^1 \frac{g(x)}{(1-x)_+} =  \int_0^1 \frac{g(x)- g(1)}{1-x} \,
\end{align}
for a smooth test function $g(x)$. We write our results in terms of Harmonic PolyLogarithms~\cite{Remiddi:1999ew}. We use the \textsc{Mathematica} package \textsc{Hpl}~\cite{Maitre:2005uu} to manipulate them. We use the standard shorthand notation
\begin{align}
  \label{eq:HPL}
  H_{a_1, \ldots, a_n} \equiv {\rm HPL}(a_1, \ldots, a_n; x) \,.
\end{align}
We have also obtained the two-loop results through $\Ord(\e^2)$, which are required for future N$^3$LO calculation. We do not show them here to save space, but instead include them as ancillary file in the arXiv submission of this paper. 

\subsection{Renormalization counter terms and zero-bin subtraction}
\label{sec:renorm-count-terms}

The bare results contain both UV and rapidity divergences to be renormalized, while the rapidity divergence is already renormalized within exponentional regularization upon the substitution $\tau \to 1/\nu$. To proceed, we first perform the usual coupling constant renormalization in $\overline{\rm MS}$ scheme,
\begin{align}
  \label{eq:coupling_ren}
  \alpha_0 \mu_0^{2 \e} ( 4\pi)^\e e^{-\e \gamma_E} = \alpha_s \mu^{2 \e} \left(
1 - \frac{\alpha_s}{4 \pi} \frac{\beta_0}{\e} + \Ord(\alpha_s^2) \right) \,,
\end{align}
where $\gamma_E = 0.5772 \ldots$ is the Euler constant, $\mu_0$ is the mass parameter in dimensional regularization, $\alpha_s = \alpha_s(\mu)$ is the renormalized coupling evaluated at the renormalization scale $\mu$. The one-loop QCD beta function for $N_f$ light flavor is given by
\begin{align}
  \label{eq:beta0}
  \beta_0 = \frac{11}{3} C_A - \frac{4}{3} T_F N_f \,.
\end{align}

Next, we perform a zero-bin subtraction~\cite{Manohar:2006nz} to remove the overlapping contributions between collinear and soft modes. For the purpose of computing perturbative matching coefficients, we can used the dimensional regularized collinear PDFs in a partonic state,
\begin{align}
\phi_{i/j}(x,\alpha_s) &= \delta_{ij} \delta(1-x) - \alspi \frac{P_{ij}^{\zero}(x)}{\e} \nn
\\
&+ \left( \alspi \right)^2
\left[ \frac{1}{2\e^2} \left( \sum_{k} P_{ik}^\zero(x) \otimes P_{kj}^\zero(x) + \beta_0 P_{ij}^\zero(x) \right) - \frac{P_{ij}^\one(x)}{2\e} \right] \,,
\label{eq:pdfct}
\end{align}
where $P_{ij}(x)$ are space-like splitting kernel, whose explicit expressions are collected in the appendix.
After this, the remaining UV divergences in the TMDPDFs can be removed by a multiplicative renormalization counter term. These steps can be summarized as
\begin{align}
  \label{eq:renbeam}
  \frac{\widetilde{{\cal B }}_{g/j}^{{\rm bare},\mu\nu}}{\mathcal{S}_{0 \rm b}} = Z_g^B {\cal B}_{g/j}^{\mu\nu} = Z_g^B \sum_i \mathcal{I}^{\mu \nu}_{g i} \otimes \phi_{i/j}  \,,
\end{align}
where $\widetilde{{\cal B }}_{g/j}^{{\rm bare},\mu\nu}$ and $\mathcal{S}_{0 \rm b}(\alpha_s)$ are the bare TMDPDFs and bare zero-bin soft function in terms of renormalized coupling $\alpha_s$, $Z_g^B$ is the multiplicative operator renormalization constant, and in the last equality we use the renormalized version of operator product expansion in Eq.~\eqref{eq:PDFOPE}.
The zero-bin soft function can be found in appendix.~\ref{sec:zerobinsoft}.

\subsection{Renormalized coefficient functions}
\label{sec:renorm-coeff-funct}

In this subsection, we present the detailed results for the renormalized matching coefficient functions through $\Ord(\alpha_s^2)$. The renormalized coefficients obey a RG equation
\begin{multline}
\frac{\df}{\df \ln\mu} \cI_{gi}^{ \mu \nu}(x,b_\perp,\mu,\nu) = 2 \left[ \Gcusp(\alsmu) \ln\frac{\nu}{xP_{+}} + \gamma^B(\alsmu) \right] \cI_{gi}^{\mu \nu}(x,b_\perp,\mu,\nu)
\\
- 2 \sum_j \cI_{gj}^{\mu \nu}(x,b_\perp,\mu,\nu) \otimes P_{ji}(x,\alsmu) \,,
\label{eq:Imu}
\end{multline}
and a rapidity evolution equation~\cite{Chiu:2012ir}
\begin{equation}
\frac{\df}{\df\ln\nu} \cI_{gi}^{\mu \nu}(x,b_\perp,\mu,\nu) = -2 \left[ \int_{\mu}^{b_0/b_T} \frac{\df\bar{\mu}}{\bar{\mu}} \Gcusp(\alpha_s(\bar{\mu})) + \gamma^R(\als(b_0/b_T)) \right] \cI_{gi}^{\mu \nu}(x,b_\perp,\mu,\nu) \, .
\label{eq:Inu}
\end{equation}
The relevant anomalous dimensions are collected in Appendix.~\ref{sec:AD}.
The tensor decomposition for the renormalized quantities is the same as given in Eq.~\eqref{eq:PDFdecomposition}. Throughout this paper, we define the pertubative expansion according to
\begin{align}
\mathcal{I}_{gi}(x, b_\perp,L_Q) = \sum_j   \left( \frac{\alpha_s}{4 \pi}\right)^j \cI_{gi}^{(j)}(x,b_\perp,L_Q) \,. 
\end{align}
Then, the renormalized scalar form factors are given by
\begin{align}
\cI^{'\one}_{gi}(x,b_\perp,L_Q) & =  I^{'(1)}_{gi} \,, \nn
\\
\cI^{'\two}_{gi}(x,b_\perp,L_Q)&=\bigg[\bigg(\beta_0-\frac{1}{2} \Gcusp_0 L_Q+\gamma_0^B\bigg)I^{'(1)}_{gi}-\sum_j I^{'(1)}_{gj}\otimes P_{ji}^\zero\bigg]L_\perp \nonumber \\ 
&+ \gamma_0^R L_Q I^{'(1)}_{gi}+I^{'(2)}_{gi}\,,\nn
 \\
\cI^\zero_{gi}(x,b_\perp,L_Q) &= \delta_{gi} \delta(1-x)  \,, \nn
\\
\cI^\one_{gi}(x,b_\perp,L_Q) &= \left( - \frac{\Gcusp_0}{2} \Lp L_Q + \gamma_0^B \Lp + \gamma_0^R L_Q \right) \delta_{gi} \delta(1-x) - P_{gi}^\zero(x) \Lp + I_{gi}^\one(x) \,, \nn
\\
\cI^\two_{gi}(x,b_\perp,L_Q) &= \bigg[ \frac{1}{8} \left( -\Gcusp_0 L_Q + 2\gamma^B_0 \right) \left( -\Gcusp_0 L_Q + 2\gamma^B_0 + 2\beta_0 \right) \Lp^2 \nn
\\
& + \left( -\frac{\Gcusp_1}{2} L_Q + \gamma^B_1 + (-\Gcusp_0 L_Q + 2\gamma^B_0 + 2\beta_0) \frac{\gamma_0^R}{2} L_Q \right) \Lp \nn
\\
& + \frac{(\gamma_0^R)^2}{2} L_Q^2 + \gamma_1^R L_Q \bigg] \, \delta_{gi} \delta(1-x) \nn
\\
&+ \bigg( \frac{1}{2} \sum_j P^\zero_{gj}(x) \otimes P^\zero_{ji}(x) + \frac{P^\zero_{gi}(x)}{2} (\Gcusp_0 L_Q - 2\gamma_0^B - \beta_0) \bigg) \Lp^2 \nn
\\
&+ \bigg[ -P^\one_{gi}(x) - P^\zero_{gi}(x) \gamma_0^R L_Q - \sum_j I^\one_{gj}(x) \otimes P^\zero_{ji}(x) \nn
\\
& + \left( -\frac{\Gcusp_0}{2} L_Q + \gamma_0^B + \beta_0 \right) I^\one_{gi}(x) \bigg] \Lp + \gamma_0^R L_Q I^\one_{gi}(x) + I^\two_{gi}(x) \,,
\end{align}
where we have shown explicitly the scale-dependent part and scale-independent part. At each order, the scale-dependent part are determined by RG equations and universal anomalous dimensions. They serve as strong check of the results from Feynman diagram calculation. The genuine new results of direct calculations are the scale-independent terms. At one loop they are given by
\begin{align}
I^{'(1)}_{gq}&=4C_F\,r_b(x)\,,\nonumber \\
I^{'(1)}_{gg}&=4C_A\,r_b(x) \,, \nn \\
I^{(1)}_{gq}&=2C_F \,x\,,\nonumber \\
I^{(1)}_{gg}&=0 \,,
\label{eq:TMDPDFoneloop}
\end{align}
where $r_b(x)$ is defined in Eq.~\eqref{eq:rb}. At two loops, we find 
\begin{align}
\label{eq:TMDPDFtwoloop}
I^{'(2)}_{gq}&=C_A C_F \bigg[ -16r_b(x)\left(-H_{1,1}+\frac{5}{6}H_{1}-H_{2}+\zeta_2\right)+16H_{0,0}-\frac{8(5x+2)}{x}H_{0}\nonumber \\
&+\frac{8(1-x)(9x-11)}{9x}\bigg] \nonumber \\&+C_F^2\bigg[-16r_b(x)H_{1,1}-8H_{0,0}-8H_{0,0}+8r_b(x)H_{1}+20H_{0}-8(1-x)\bigg]\nonumber \\
&+N_f C_F T_F\bigg[-\frac{64}{9}r_b(x)+\frac{32}{3}r_b(x) H_{1} \bigg]\,, \nn \\
I^{'(2)}_{gg}&=C_A^2\bigg[ 16r_b(x)\left(-\zeta_2+H_{2}\right)+16H_{0,0}-\frac{4(19x+12)}{3x}H_{0}+\frac{4(11x^3-39x^2+62x-37)}{9x}\bigg] \nonumber \\
&+N_f C_A T_F\bigg[\frac{8(x^3+3x^2+16x-17)}{9x}-\frac{16}{3}H_{0}\bigg] +N_f C_F T_F\bigg[ -16H_{0,0}+\frac{16(1-x)^3}{x}\bigg]\,, \nn \\
I^{(2)}_{gq}&=C_A C_F \bigg[ \frac{2}{3}(8x^2+9x+36)H_{0,0}-\frac{2(5x^2-22x+22)}{3x}H_{1,1}+\frac{8(x^2+2)}{x}H_{2,0}\nonumber \\
&-\frac{4(4x^3-9x^2+24x-22)}{3x}H_{1,0}+4xH_{-1,0}-4(x+2)H_{0,0,0}-\frac{2}{9}(88x^2-6x+249)H_{0}\nonumber \\
&+\frac{2(43x^2-152x+152)}{9x}H_{1}+4xH_{2}+\frac{8(3x^2-7x+6)}{x}\zeta_3+\frac{8(1-x)(2x^2-x+11)}{3x}\zeta_2\nonumber \\
&+\frac{4(152x^3-268x^2+791x-790)}{27x}+4p_{gq}(x)\left( H_{1,2}+H_{2,1}-H_{1,0,0}
+H_{1,1,0}+H_{1,1,1}\right) \nonumber \\
&-4p_{gq}(-x)\left(-2 H_{-2,0}+2H_{-1,-1,0}-H_{-1,0,0}+\zeta_2 H_{-1}\right)
\bigg] \nonumber \\
&+C_F^2\bigg[ \frac{2(x^2-6x+6)}{x}H_{1,1}-\frac{4(x^2-2x+2)}{x}H_{1,1,1}+(-3x-4)H_{0,0}+2(2-x)H_{0,0,0}\nonumber \\
&-\frac{2(5x^2-16x+16)}{x}H_{1}+5(x-3)H_{0}-x+10\bigg]\nonumber \\
&+N_f C_F T_F\bigg[\frac{8(x^2-2x+2)}{3x}H_{1,1}-\frac{16(x^2-5x+5)}{9x}H_{1}+\frac{8(13x^2-56x+56)}{27x}\bigg] \,, \nn \\
I^{(2)}_{gg}&=C_A^2\bigg[  \left( 28 \zeta_3 - \frac{808}{27} \right) \frac{1}{(1-x)_+}+\frac{4(835x^3-760x^2+926x-790)}{27x} -\frac{2}{3}x H_{1} \nonumber \\
&+\frac{2}{3}(44x^2-11x+25)H_{0,0}-\frac{16(x^4-x^2-1)}{(1-x)x(x+1)}H_{2,0}\nonumber \\
&+\frac{1}{9}(-536x^2-149x-701)H_{0}+ \frac{8(1-x)(11x^2-x+11)}{3x}\left(\zeta_2+H_{1,0}\right)\nonumber \\
&+\frac{4(14x^5-12x^4+14x^3+5x^2-21x+12)}{(1-x)x(x+1)}\zeta_3 -\frac{8(x^2-x-1)^2}{(1-x)(x+1)}H_{0,0,0}\nonumber \\
&-8(-2H_{-2,0}+2H_{-1,-1,0}-H_{-1,0,0}+\zeta_2H_{-1})p_{gg}(-x)\nonumber \\
&-8(-H_{1,2}-H_{2,1}+H_{1,0,0}-H_{1,1,0})p_{gg}(x)
\bigg]\nonumber \\
&+N_f C_F T_F\bigg[ 4(x+3)H_{0,0}+8(x+1)H_{0,0,0}+24(x+1)H_{0}-\frac{8(1-x)(x^2-23x+1)}{3x}\bigg] \nn \\
&+N_f C_A T_F\bigg[  \frac{224}{27} \frac{1}{(1-x)_+} -\frac{4(139x^3-110x^2+166x-121)}{27x}\nonumber \\
&+\frac{8}{3}(x+1)H_{0,0}+\frac{4}{9}(10x+13)H_{0}+\frac{4}{3}xH_{1}\bigg] \,.
\end{align}
The two-loop scalar form factors corresponding to the first tensor structure, $I_{gi}^{(2)}$ , have been computed for a while~\cite{Gehrmann:2014yya,Echevarria:2016scs}. Results for the second tensor structure $I^{'(2)}_{gi}$, also known as the linearly polarized contribution,  appeared very recently, using a different rapidity regulator~\cite{Gutierrez-Reyes:2019rug}. Although the calculations are performed with different rapidity regulators, the results can be compared by constructing a rapidity-divergence free combination of TMDPDFs. In our case, this can be done by multiplying the renormalized coefficient functions with the square root of the TMD soft function,
\begin{align}
  \label{eq:compareTMDPDF}
\widetilde{{\cal I}}_{gi}^{\mu\nu}(x,b_\perp,Q,\mu) =  {\cal I}_{gi}^{\mu\nu}(x, b_\perp, L_Q) \sqrt{{\cal S}_{gg}(b_\perp,\mu,\nu)}  \,,
\end{align}
for $i = q, g$,  and then expand in $\alpha_s$. The renormalized TMD soft function can be found in Ref.~\cite{Li:2016ctv}. The explicit two-loop results for the renormalized soft function are collected in Eq.~\eqref{eq:soft} of appendix.~\ref{sec:reTMDsoft}.
We find that for the first tensor structure, our two-loop results are in full agreement with those in the literature~\cite{Gehrmann:2012ze,Echevarria:2016scs}. However, we find substantial difference for the two-loop linearly polarized results with those presented in Ref.~\cite{Gutierrez-Reyes:2019rug}. We note that the two-loop results in Ref.~~\cite{Gutierrez-Reyes:2019rug} contain transcendental functions and zeta value up to weight $3$, namely ${\rm Li}_3$ and $\zeta_3$, while in our results, $I^{'(2)}_{gg}$ and  $I^{'(2)}_{gq}$ in Eq.~\eqref{eq:TMDPDFtwoloop}, only weight $2$ functions and zeta values are presented.  Since the linearly polarized contribution first appears at one loop with rational functions only, our results are in agreement with the expectation that transcendental weight only increases by $2$ at each loop order. In the next subsection, we present a ${\cal N}=1$ supersymmetry sum-rule for the linearly polarized gluon contribution, which provides further check to our results. 

\subsection{${\cal N}=1$ supersymmetry sum rule for the linearly polarized gluon contribution}
\label{sec:cal-n=1-supersymm}

Our explicit two-loop calculation reveals an interesting momentum conservation sum rule for the linearly polarized gluon contribution in the ${\cal N}=1$ supersymmetric limit. This limit is obtained from our results by setting $C_A = C_F = N_f$ and using $T_F = 1/2$. The sum rule is then written as
\begin{align}
  \label{eq:lpsumrule}
  \int_0^1 dx\, x \Big( {\cal I}^{'}_{gg}(x, b_\perp, L_Q) - {\cal I}^{'}_{gq}(x, b_\perp, L_Q)  \Big) \Big|_{C_F = C_A = N_f} = 0 \,.
\end{align}
At one-loop, the sum rule is satisfied trivially, since $I^{'(1)}_{qg}/C_F =  I^{'(1)}_{gg}/C_A = 4 r_b(x)$, see Eq.~\eqref{eq:TMDPDFoneloop}. At two loops, we find
\begin{align}
  \label{eq:twoloopsumrule}
  \Big( {\cal I}^{'(2)}_{gg}(x, b_\perp, L_Q) - {\cal I}^{'(2)}_{gq}(x, b_\perp, L_Q)  \Big) \Big|_{C_F = C_A = N_f} = - \frac{8}{3x}  C_A^2 \bigg( 1 +x  - 3 x^2 + x^3  + 3 x \ln(x) \bigg) \,,
\end{align}
substituting this into Eq.~\eqref{eq:lpsumrule} we indeed get zero. 

While it is well-known that the splitting functions obey momentum conservation sum rule, in general the sum rule breaks down for the matching coefficient functions, which are cross-section level quantities. Therefore, the sum rule in Eq.~\eqref{eq:lpsumrule} is somewhat surprising. It would be interesting to see if it continues to hold at three loops. It would also be interesting to have a structural understanding of it. Since the sum rule is nontrivial, it also provides strong check to the two-loop results for the linearly polarized gluon contribution presented in this paper. 

\section{Gluon TMDFFs}
\label{sec:gluon-tmdff}

To specify the definition for the gluon TMDFFs, it is necessary to specify a reference frame first. In the hadron frame, where the detected hadron has zero transverse momentum, the gluon TMDFFs can be defined as
\begin{align}
  \label{eq:FF_hadron_Frame}
  {\cal D}_{N/g}^{\rm bare,\mu\nu} (z, b_\perp) = 
- \frac{P_+}{z^2} \sum_X \int \frac{db_-}{4 \pi} 
e^{i x b_- P_+/2} \langle 0 |
{\cal A}_{n\perp}^{a,\mu}(0, b_-, b_\perp) | N(P), X \rangle
\langle N(P), X | {\cal A}_{n\perp}^{a,\nu}(0) | 0 \rangle \,,
\end{align}
where $N(P)$ has zero transverse momentum. In actual calculation, it's also convenient to define the fragmentation functions in the parton frame, where the parton which initiates the fragmentation has zero transverse momentum. The parton frame TMDFFs are related to the hadron frame ones by
\begin{align}
  \label{eq:partontohadron}
  {\cal F}_{N/g}^{{\rm bare},\mu\nu} (z, b_\perp/z) = z^{2 - 2 \e} {\cal D}_{N/g}^{{\rm bare},\mu\nu} (z, b_\perp)  \,,
\end{align}
where we denote the bare TMDFFs in the parton frame by $ {\cal F}_{N/g}^{{\rm bare},\mu\nu} (z, b_\perp/z)$. The parton frame has the advantage that the renormalization counter terms are slightly simpler compared with the hadron frame, i.e. 
\begin{align}
\mathcal{F}^{\text{bare}, \mu \nu}_{N/g}(z, b_\perp/z,\nu) 
&\,= Z^B_g(b_\perp,\mu,\nu) \, \mathcal{F}^{\mu \nu}_{N/g}(z,b_\perp/z,\mu,\nu) 
\nonumber
\\
&\,= Z^B_g(b_\perp,\mu,\nu) \sum_i  d_{N/i}(z,\mu)  \otimes \mathcal{C}^{ \mu \nu}_{ig}(z,b_\perp/z,\mu,\nu) + \mathcal{O}(b_T^2\Lambda^2_{\text{QCD}}) \, .
\end{align}
We refer to Refs.~\cite{Collins:2011zzd,Luo:2019hmp} for more detailed discussion on the difference of the reference frames.

To compute the gluon TMDFFs, we can use the methods as in the calculation of TMDPDFs. Alternatively, we can exploit crossing symmetry to obtain the results from TMDPDFs. We have performed calculation in both ways, and find the same results. In the next subsection, we give some details on the calculation based on crossing relation.

\subsection{Bare gluon TMDFFs from crossing}
\label{sec:bare-results-1}

To explore the crossing symmetry between the TMDPDFs and TMDFFs, we write down their definitions in momentum space. For simplicity we first consider the $g \to g$ case,
\begin{multline}
\mathcal{B}^{\text{bare}, \mu \nu}_{g/g}(x,b_\perp,p_1,\nu) = \lim_{\tau \to 0} \frac{1}{x \bar{n}\cdot p_1} \int d^dk \, e^{-b_0 \tau k^0 + i \bp \cdot \vec{k}_{T}}  \, \delta( \bar{n} \cdot k  - (1-x) \bar{n} \cdot p_1)
\\ 
\times \prod \int \frac{d^dl_j}{(2\pi)^d} \prod \int \frac{d^dk_i}{(2\pi)^d}  \, (2\pi) \delta_{+}(k_i^2) \, \delta^{(d)}\bigg(k-\sum_m k_m \bigg) \, |\mathcal{M}^{B \,\mu \nu}_{gg}(p_1,l_j,k_i)|^2 \bigg|_{\tau=1/\nu} \,,
\label{eq:TMDPDFdef}
\end{multline}
for TMDPDFs. In Eq.~\eqref{eq:TMDPDFdef}, $|\mathcal{M}^{B \,\mu \nu}_{gg}(p_1,l_j,k_i)|^2$ is the squared amplitudes for space-like $g\to g$ splitting, with multiple real radiations~($k_i$) or virtual momentum exchange~($l_j$), and $p_1$ is the momenta of the initial-state gluon entering hard scattering. For TMDFFs in hadron frame, we have 
\begin{multline}
\mathcal{D}^{\text{bare},\mu \nu}_{g/g}(z,b_\perp,p_2,\nu) = \lim_{\tau \to 0} \frac{1}{\bar{n} \cdot p_2} \int d^dk \, e^{-b_0 \tau k^0 + i \bp \cdot \vec{k}_{T}}  \, \delta( \bar{n} \cdot k - (1/z-1) \bar{n} \cdot p_2)
\\ 
\times \prod \int \frac{d^dl_j}{(2\pi)^d} \prod \int \frac{d^dk_i}{(2\pi)^d}  \, (2\pi) \delta_{+}(k_i^2) \, \delta^{(d)}\bigg(k-\sum_m k_m \bigg) \, |\mathcal{M}^{F \, \mu \nu}_{gg}(p_2,l_j,k_i)|^2 \bigg|_{\tau=1/\nu} \,,
\end{multline}
where $|\mathcal{M}^{F \,\mu \nu}_{gg}(p_2,l_j,k_i)|^2$ is the squared amplitudes for time-like $g\to g$ splitting,
 and $p_2$ is the momenta of the final-state detected gluon, which has zero transverse momentum in hadron frame. The squared amplitudes in the integrand for TMDPDFs and TMDFFs are related through the following crossing relation, 
\begin{align}
|\mathcal{M}^{F\, \mu \nu}_{gg}(p_2,l_j,k_i)|^2 =|\mathcal{M}^{B \, \mu \nu}_{gg}(-p_2,l_j,k_i)|^2  \,.
\end{align}
It is not difficult to see that
\begin{align}
\mathcal{B}^{\text{bare}, \mu \nu}_{g/g}&(\frac{1}{z},b_\perp,-p_2,\nu)= \lim_{\tau \to 0} \frac{z}{- \bar{n}\cdot p_2} \int d^dk \, e^{-b_0 \tau k^0 + i \bp \cdot \vec{k}_{T}}  \, \delta( \bar{n} \cdot k  + (1-1/z) \bar{n} \cdot p_2)
\nonumber \\  
&\times \prod \int \frac{d^dl_j}{(2\pi)^d} \prod \int \frac{d^dk_i}{(2\pi)^d}  \, (2\pi) \delta_{+}(k_i^2) \, \delta^{(d)}\bigg(k-\sum_m k_m \bigg) \, |\mathcal{M}^{F \mu \nu}_{gg}(p_2,l_j,k_i)|^2 \bigg|_{\tau=1/\nu}  \nonumber \\
& = -z \mathcal{D}^{\text{bare},\mu \nu}_{g/g}(z,b_\perp,p_2,\nu) \,,
\end{align}
that is,
\begin{align}
\mathcal{D}^{\text{bare},\mu \nu}_{g/g}(z,b_\perp,Q_2,\nu) = -\frac{1}{z} \mathcal{B}^{\text{bare}, \mu \nu}_{g/g}&(x,b_\perp,Q_1,\nu)|_{x \to \frac{1}{z}, Q_1 \to -Q_2} \,,
\end{align}
where we have defined
\begin{align}
Q_1 = x  \, \bar{n}  \cdot p_1 \,, \quad Q_2 = \frac{\bar{n} \cdot p_2}{z} \,.
\end{align}
The rule $Q_1 \to -Q_2$ is important for resolving the ambiguity from analytic continuation of $\ln(1-x)$~(see also discussions in \cite{Stratmann:1996hn, Blumlein:2000wh, Mitov:2006ic, Moch:2007tx, Almasy:2011eq} for analytic continuation of splitting funtions), which can have two possibilities,
\begin{align}
\label{rulexEq}
\ln(1-x)  \to \ln(1-x) - \ln(x) + \kappa \, i \pi   \quad \text{with} \quad \kappa = 0  \quad \text{or} \quad 1 \,.
\end{align}
We argue that one should set $\kappa = 0 $ for $\ln(1-x)$ terms originate from rapidity divergences, and $\kappa = 1$ for $\ln(1-x)$ terms originate from virtual corrections. For virtual corrections, the analytical continuation is unambiguous, since it is determined by Feynman's $i \varepsilon$ prescription. To understand the prescription for the $\ln(1-x)$ from rapidity divergences, we first introduce the following dimensionless variables,
\begin{align}
1- x= \frac{\bar{n} \cdot k}{\bar{n} \cdot p_1}\,, \quad   \quad 1-y_1 = \frac{k^2  \, \bar{n} \cdot p_1}{2 p_1 \cdot k \, \bar{n} \cdot k}\,, \nonumber \\
\frac{1-z}{z } = \frac{\bar{n} \cdot k}{\bar{n} \cdot p_2} \,, \quad   \quad 1-y_2 = \frac{k^2  \, \bar{n} \cdot p_2}{2 p_2 \cdot k  \, \bar{n} \cdot k} \,.
\end{align}
Using these variables, the exponential regulator in TMDPDFs and TMDFFs can be separately written as~\cite{Li:2016axz,Luo:2019hmp}
\begin{align}
\exp(-2 \tau k^0) &\, =  \exp\left( \frac{-\tau x k_T^2 }{y_1 (1-x) Q_1} -\tau (1-x) \frac{Q_1}{x} \right) \,, \nonumber \\
\exp(-2 \tau k^0) &\, = \exp\left( \frac{-\tau k_T^2 }{ y_2 (1-z) Q_2} - \tau (1-z) Q_2\right)\,,
\end{align}
where we have used that $2 k^0 = \bar{n} \cdot k + n \cdot k$.
These two expressions can be exactly related to each other through the following rules,
\begin{align}
p_1 \to -p_2\,, \quad  k \to k\,, \quad  x \to \frac{1}{z}\,, \quad Q_1 \to -Q_2 \,.
\end{align}
The exponential regulator regularizes the rapidity divergences in the phase space integral only. Therefore, the $\ln(1-x)$ terms that originate from rapidity divergence should not develop imaginary part under the replacements $ x \to 1/z$,  $Q_1 \to -Q_2$. This is indeed the case since the logarithm always appear in a specific combination
\begin{align}
\ln\left(\frac{ \tau x}{ (1-x) Q_1} \right)\,.
\end{align}

In practical calculation, one prescription to resolve the ambiguity in Eq.~\eqref{rulexEq} is to use the the following crossing rules
\begin{align}
\label{ruleQEq}
\ln(Q_1) \to \ln(Q_2) - i \pi \,, \nonumber \\ 
\ln(1-x)  \to \ln(1-z) - \ln(z) + i \pi \,.  
\end{align}
Moreover, taking into account the relation between parton frame and hadron frame, which is given in Eq.~\eqref{eq:partontohadron},
and the color and spin factors, the analytic continuations from TMDPDFs to TMDFFs reads
\begin{align}
\mathcal{F}^{\text{bare},\mu \nu}_{g/g}(z, b_\perp/z,Q_2,\nu) = -z^{1- 2 \epsilon} \mathcal{B}^{\text{bare}, \mu \nu}_{g/g}&(x,b_\perp,Q_1,\nu)|_{x \to \frac{1}{z}, L_{Q_1} \to L_{Q_2} - 2 i \pi} \,, \nonumber \\
\mathcal{F}^{\text{bare},\mu \nu}_{q/g}(z, b_\perp/z,Q_2,\nu) = \frac{ z^{1- 2 \epsilon} T_F}{C_F(1-\epsilon)}\mathcal{B}^{\text{bare}, \mu \nu}_{g/q}&(x,b_\perp,Q_1,\nu)|_{x \to \frac{1}{z},L_{Q_1} \to L_{Q_2} - 2 i \pi} \,,
\label{eq:crossing}
\end{align}
where
\begin{align}
L_{Q_1} = 2 \ln \left( \frac{Q_1}{\nu} \right) \,, \quad  L_{Q_2} = 2 \ln \left( \frac{Q_2}{\nu} \right) \,. 
\end{align} 
In the final step, we take the real part in Eq.~\eqref{eq:crossing}. Note that the analytic continuations in Eq.~\eqref{eq:crossing} are valid in the region
\begin{align}
0 <x <1 \,, \quad  0<z<1 \,.
\end{align} 
The contributions from the end point, $\delta(1-x)$ and $\delta(1-z)$, are invariant under crossing.  

The analytic continuations in Eq.~\eqref{eq:crossing} can be also used to extract time-like splitting functions from space-like ones. It is similar to the analytic continuations of splitting function performed in Refs.~\cite{ Mitov:2006ic, Moch:2007tx}. There the calculations start from the crossing of unrenormalized structure functions and work well for two-loop splitting functions. At three loops, the direct analytic continuations cause some issues~(see \cite{ Mitov:2006ic, Moch:2007tx} for details), so it is expected our procedure presented here may also cause similar issues at three loops. However, this is beyond the scope of this paper.

\subsection{Renormalization counter terms and zero-bin subtraction}
\label{sec:renorm-count-terms-1}

The renormalization of the TMDFFs are similar to the renormalization of the TMDPDFs. In the parton frame, we use the following dimensional regularized collinear FFs for the counter terms:
\begin{align}
d_{i/j}(z,\mu) &= \delta_{ij} \delta(1-z) - \alsmupi \frac{P_{ij}^{T\zero}(z)}{\e} \nn
\\
&+ \left( \alsmupi \right)^2
\left[ \frac{1}{2\e^2} \left( \sum_{k} P_{ik}^{T \zero}(z) \otimes P_{kj}^{T \zero}(z) + \beta_0 P_{ij}^{T \zero}(z) \right) - \frac{P_{ij}^{T \one}(z)}{2\e} \right] \,,
\end{align}
where $P_{ij}^T(z)$ are the time-like splitting kernel, whose explicit expression through two loops can be found in appendix.~\ref{sec:timelikesplitting}.
In hadron frame, the counter terms from collinear FFs involve additional factor of $z^{-2 \e}$~\cite{Echevarria:2016scs}, which arises from the phase space factor. The zero-bin subtraction also follows closely the TMDPDFs.

\subsection{Renormalized coefficient functions}
\label{sec:renorm-coeff-funct-1}

In this subsection, we present the renormalized coefficient functions through two loops for both polarizations. We separate the results into scale-independent part and scale-dependent part. The scale-dependent part is determined by the the RG equation
\begin{multline}
\frac{\df}{\df \ln\mu} \cC^{ \mu \nu}_{ig}(z,b_\perp/z,\mu,\nu) = 2 \left[ \Gcusp(\alsmu) \ln\frac{z\nu}{P_{+}} + \gamma^B(\alsmu) \right] \cC^{\mu \nu}_{ig}(z,b_\perp/z,\mu,\nu)
\\
- 2 \sum_j P^T_{ij}(z,\alsmu) \otimes \cC^{\mu \nu}_{jg}(z,b_\perp/z,\mu,\nu) \, ,
\label{eq:Cmu}
\end{multline}
and rapidity evolution equation 
\begin{equation}
\frac{\df}{\df\ln\nu} \cC^{\mu \nu}_{ig}(z,b_\perp/z,\mu,\nu) = -2 \left[ \int_{\mu}^{b_0/b_T} \frac{\df\bar{\mu}}{\bar{\mu}} \Gcusp(\alpha_s(\bar{\mu})) + \gamma^R(\als(b_0/b_T)) \right] \cC^{\mu \nu}_{ig}(z,b_\perp/z,\mu,\nu) \,.
\label{eq:Cnu}
\end{equation}
The two polarization form factors can be extracted through
\begin{align}
\label{eq:FFProjection}
\mathcal{C}_{ig}(z, b_\perp/z,L_Q) &= g_\perp^{ \mu \nu} {\cal C}_{ig}^{ \mu\nu} (z, b_\perp/z, \mu, \nu)\,, \nonumber \\
\mathcal{C}'_{ig}(z, b_\perp/z,L_Q) &=  \frac{1}{d-3} \left[  g_\perp^{\mu \nu} + \left(   d-2\right) \frac{b_\perp^\mu b_\perp^\nu}{b_T^2} \right] {\cal C}_{ig}^{\mu\nu} (z, b_\perp/z,\mu,\nu) \,.
\end{align}
Compared with Eq.~\eqref{eq:Ldefinition}, for TMDFFs we use a slightly different definition for $L_Q$,
\begin{align}
L_Q = 2 \ln \frac{P_+}{ z \, \nu} \,. 
\end{align}
The explicit solutions to both equations through to two loops are given by 
\begin{align}
\cC^{'\one}_{ig}(z,b_\perp/z,L_Q) & =  C^{'(1)}_{ig}\, , \nn
\\
\cC^{'\two}_{ig}(z,b_\perp/z,L_Q) &=\bigg[\bigg(\beta_0-\frac{1}{2} \Gcusp_0 L_Q+\gamma_0^B\bigg)C^{'(1)}_{ig}-\sum_j P_{ij}^{T\zero} \otimes C^{'(1)}_{jg}\bigg]L_\perp \nonumber \\ 
&+ \gamma_0^R L_Q C^{'(1)}_{ig}+C^{'(2)}_{ig}\,,\nn
 \\
\cC^\zero_{ig}(z,b_\perp/z,L_Q) &= \delta_{ig} \delta(1-z) \, ,\nn
\\
\cC^\one_{ig}(z,b_\perp/z,L_Q) &= \left( - \frac{\Gcusp_0}{2} \Lp L_Q + \gamma_0^B \Lp + \gamma_0^R L_Q \right) \delta_{ig} \delta(1-z) - P_{ig}^{T\zero}(z) \Lp + C_{ig}^\one(z) \, , \nn
\\
\cC^\two_{ig}(z,b_\perp/z,L_Q) &= \bigg[ \frac{1}{8} \left( -\Gcusp_0 L_Q + 2\gamma^B_0 \right) \left( -\Gcusp_0 L_Q + 2\gamma^B_0 + 2\beta_0 \right) \Lp^2 \nn
\\
&\hspace{1em} + \left( -\frac{\Gcusp_1}{2} L_Q + \gamma^B_1 + (-\Gcusp_0 L_Q + 2\gamma^B_0 + 2\beta_0) \frac{\gamma_0^R}{2} L_Q \right) \Lp \nn
\\
&\hspace{1em} + \frac{(\gamma_0^R)^2}{2} L_Q^2 + \gamma_1^R L_Q \bigg] \, \delta_{ig} \delta(1-z) \nn
\\
&+ \bigg( \frac{1}{2} \sum_j P^{T\zero}_{ij}(z) \otimes P^{T\zero}_{jg}(z) + \frac{P^{T\zero}_{ig}(z)}{2} (\Gcusp_0 L_Q - 2\gamma_0^B - \beta_0) \bigg) \Lp^2 \nn
\\
&+ \bigg[ -P^{T\one}_{ig}(z) - P^{T\zero}_{ig}(z) \gamma_0^R L_Q - \sum_j P^{T\zero}_{ij}(z) \otimes C^\one_{jg}(z) \nn
\\
&\hspace{1em} + \left( -\frac{\Gcusp_0}{2} L_Q + \gamma_0^B + \beta_0 \right) C^\one_{ig}(z) \bigg] \Lp + \gamma_0^R L_Q C^\one_{ig}(z) + C^\two_{ig}(z) \,.
\label{eq:RGexpC}
\end{align} 
Our explicit diagrammatic calculation or analytical continuation reproduces all the scale-dependent terms, which serve as a first check to the calculation. The genuine new results at each order are the scale independent part. At one loop they are given by
\begin{align}
C^{'(1)}_{qg}&=-4T_F\,r_f(z)\,,\nonumber \\
C^{'(1)}_{gg}&=4C_A\,r_f(z)\,, \nn \\
C^{(1)}_{qg}&=4 T_F \bigg(p_{qg}(z) H_{0}+r_f(z)\bigg)\,,\nonumber \\
C^{(1)}_{gg}&=8C_A\,p_{gg} (z) H_{0}\,,
\end{align}
where we have defined 
\begin{align}
r_f(z) = (1-z)z \,. 
\end{align}
The two-loop results are
\begin{align}
C^{'(2)}_{qg}&=C_A T_F \bigg[ -r_f(z)\bigg(16H_{1,0}+16H_{1,1}-\frac{40}{3}H_{1}-32\zeta_2+\frac{200}{9}\bigg)+16z(2z+3)H_{0,0}\nonumber \\
&-\frac{8}{3}(17z^2-26z-3)H_{0}\bigg]\nonumber \\
&+C_F T_F \bigg[ -r_f(z)\bigg(-48H_{1,0}-16H_{1,1}+8H_{1}-16H_{2}+16\zeta_2\bigg)-24z(2z-1)H_{0,0}\nonumber \\
&-4(2-z)(2z+1)H_{0}+4(1-z)(5z-1)\bigg] \nonumber \\
&+N_f T_F^2 \bigg[ -r_f(z)\bigg(\frac{32}{3}H_{0}+\frac{32}{3}H_{1}-\frac{64}{9}\bigg)\bigg] \,,\nonumber \\
C^{'(2)}_{gg}&=C_A^2\bigg[ -r_f(z)\bigg(32H_{1,0}+16H_{2}+16\zeta_2\bigg)+16(z-6)zH_{0,0}+\frac{4(12z^3-67z^2-12z+4)}{3z}H_{0}\nonumber \\
&+\frac{4(99z^3-80z^2-15z-1)}{9z}\bigg] \nonumber \\
&+N_f C_F T_F\bigg[ 48zH_{0,0}+\frac{16(9z^2+6z-2)}{3z}H_{0}-\frac{16(1-z)(28z^2+10z+1)}{9z}\bigg]\nonumber \\
&+N_f C_A T_F\bigg[ \frac{8(17z^3-16z^2-3z-1)}{9z}-\frac{16}{3}zH_{0}\bigg] \,, \nn \\
C^{(2)}_{qg}&= C_A T_F \bigg[ 8p_{qg}(-z)\bigg(-H_{-2,0}-H_{-1,-1,0}-\frac{3}{2}H_{-1,0,0}-\frac{1}{2}\zeta_2 H_{-1}\bigg)\nonumber \\
&+8 p_{qg}(z)\bigg(H_{1,2}+H_{2,1}+H_{1,0,0}+H_{1,1,0}+\frac{1}{2}H_{1,1,1}-\frac{5}{2}\zeta_2H_{0}-\frac{3}{2}\zeta_2H_{1}-\frac{5}{2}H_{3}\bigg)\nonumber \\
&-\frac{2}{3}(34z^2-34z+11)H_{1,1}-4(6z^2-10z+3)H_{2,0}-\frac{4(59z^3-35z^2+22z-16)}{3z}H_{0,0}\nonumber \\
&-\frac{2(86z^3-90z^2+45z-8)}{3z}H_{1,0}+8(z+1)zH_{-1,0}+4(62z+11)H_{0,0,0}\nonumber \\
&+\frac{4}{9}(91z^2-91z+38)H_{1}-2(14z^2-14z+11)H_{2}+\frac{2(38z^3-699z^2-165z+24)}{9z}H_{0}\nonumber \\
&\frac{2(70z^3-54z^2+21z+8)}{3z}\zeta_2+\frac{2(774z^3+139z^2-1223z-148)}{27z}+24z\zeta_3
\bigg]\nonumber \\
&+C_F T_F\bigg[ 8p_{qg}(z)\bigg(-\frac{3}{2}H_{1,2}-\frac{9}{2}H_{2,0}-\frac{3}{2}H_{2,1}-\frac{9}{2}H_{1,0,0}-\frac{3}{2}H_{1,1,0}-\frac{1}{2}H_{1,1,1}+\frac{1}{2}\zeta_2H_{0}\nonumber \\
&+\frac{3}{2}\zeta_2H_{1}-\frac{1}{2}H_{3}-4\zeta_3\bigg)+(68z^2-52z-7)H_{0,0}+6(10z^2-10z+3)H_{1,0}\nonumber \\
&+2(10z^2-10z+3)H_{1,1}-22(4z^2-2z+1)H_{0,0,0}+(-76z^2+73z-8)H_{0}\nonumber \\
&-4(9z^2-9z+4)H_{1}+2(14z^2-14z+9)H_{2}+2(-14z^2+14z-9)\zeta_2+56z^2-101z+63\bigg]\nonumber \\
&+N_f T_F^2\bigg[ 8p_{qg}(z)\bigg(\frac{1}{3}H_{0,0}+H_{1,0}+\frac{1}{3}H_{1,1}+H_{2}-\zeta_2\bigg)-\frac{8}{3}(8z^2-8z+5)H_{0}\nonumber \\
&-\frac{8}{9}(16z^2-16z+5)H_{1}+\frac{32}{27}(17z^2-17z+7)\bigg]\,, \nn \\
C^{(2)}_{gg}&=C_A^2\bigg[ \left( 28 \zeta_3 - \frac{808}{27} \right) \frac{1}{(1-z)_+}  -\frac{4(16z^5-22z^4+16z^3+29z^2-9z+22)}{(1-z)z(z+1)}\zeta_3\nonumber \\
&-\frac{4(862z^3-248z^2+10z-817)}{27z}-\frac{8(1-z)(11z^2-z+11)}{3z}H_{1,0}\nonumber \\
&+\frac{2(129z^3-144z^2+279z-220)}{3(1-z)z}H_{0,0}-\frac{16(6z^5-7z^4+6z^3+7z^2-6z+7)}{(1-z)z(z+1)}H_{2,0}\nonumber \\
&-\frac{1072z^4-407z^3+726z^2-1123z-536}{9(1-z)z}H_{0}+\frac{2}{3}H_{1}-\frac{8(1-z)(11z^2-z+11)}{3z}\zeta_2\nonumber \\
&-\frac{8(11z^5-42z^4-11z^3+42z^2+11z+20)}{(1-z)z(z+1)}H_{0,0,0}-8p_{gg}(-z)\bigg(2H_{-2,0}+2H_{-1,-1,0}
\nonumber \\
&+3H_{-1,0,0}+\zeta_2H_{-1}\bigg)-8p_{gg}(z)\bigg(H_{1,2}+H_{2,1}+7H_{1,0,0}+H_{1,1,0}+4\zeta_2H_{0}+6H_{3}\bigg)\bigg]\nonumber \\
&+N_f C_A T_F\bigg[ \frac{224}{27} \frac{1}{(1-z)_+} -\frac{4(121z^3-110z^2+166z-139)}{27z}\nonumber \\
&-\frac{8(4z^4-13z^3+12z^2-3z+4)}{3(1-z)z}H_{0,0}-\frac{4(46z^4-71z^3+90z^2-91z+46)}{9(1-z)z}H_{0}-\frac{4}{3}H_{1}\bigg]\nonumber \\
&+N_f C_F T_F\bigg[ \frac{4(16z^3+15z^2+21z+16)}{3z}H_{0,0}+88(z+1)H_{0,0,0}\nonumber \\
&-\frac{8(82z^3+81z^2+135z-6)}{9z}H_{0}-\frac{8(1-z)(301z^2+409z+139)}{27z}\bigg]\,.
\end{align}
We note that for the unpolarized coefficients, $C_{ig}$, the two-loop results have been given in Ref.~\cite{Echevarria:2016scs}. We find almost full agreement with Ref.~\cite{Echevarria:2016scs}, except a $C_A^2 \pi^4 \delta(1-z)$ term in $C_{gg}^\two$. This discrepancy is similar to a discrepancy found previously~\cite{Luo:2019hmp} for the quark TMDFFs, $C_{qq}^\two$. Like for quark TMDFFs, we have computed the gluon jet function for the EEC in the back-to-back limit and tested with known constraints to verify our results. We shall present the details of this check in the next subsection.

Our two-loop results for the linearly polarized coefficients are new. Similar to the TMDPDFs, the two-loop linearly polarized contributions involve only weight $2$ functions and zeta values. Furthermore, they obey a ${\cal N}=1$ supersymmetric momentum-conservation sum rule,
\begin{align}
  \label{eq:FFsumrule}
  \int_0^1 dz\, z \Big(C^{'}_{gg}(z, b_\perp, L_Q) + C^{'}_{qg}(z, b_\perp, L_Q)\Big) \Big|_{C_F = C_A = N_f} = 0 \,.
\end{align}
We note that there is a change of sign in the sum rule for $C'_{qg}$ compared with the corresponding sum rule for TMDPDFs in Eq.~\eqref{eq:lpsumrule}.

\section{Gluon jet funcion for the EEC in the back-to-back limit}
\label{sec:gluon-jet-funcion}

The gluon TMDFFs can be used to calculate the gluon jet function for the back-to-back resummation of the EEC in the Higgs gluonic decay~\cite{Luo:2019nig}. Such jet functions also appear in the back-to-back resummation for TEEC~\cite{Gao:2019ojf}. The gluon jet function can be expanded as
\begin{align}
  \label{eq:gluon_expanded}
  J^g(b_\perp,\mu, \nu, \alpha_s) = 1 + \sum_{n=1} \left( \frac{\alpha_s}{4 \pi} \right)^n J_n^g( b_\perp, \mu, \nu) \,.
\end{align}
The expansion coefficients through two loops are given by the second moment of the gluon TMDFFs,
\begin{align}
  \label{eq:jetfunctions}
  J_1^g =&\  \int_0^1 dz\, z \Big( {C}_{gg}^{(1)} + 2 N_f {C}_{qg}^{(1)}
+
 {C}_{gg}^{'(1)} + 2 N_f {C}_{qg}^{'(1)}\Big)
\nn\\
=&\
C_A \left(-8 \zeta _2-2 L_\perp L_Q+\frac{11 L_\perp}{3}+\frac{71}{18}\right)+N_f
   \left(-\frac{2 L_\perp}{3}-\frac{11}{18}\right) \,,
\nn\\
  J_2^g =&\  \int_0^1 dz\, z \Big( {C}_{gg}^{(2)} + 2 N_f {C}_{qg}^{(2)}
+
 {C}_{gg}^{'(2)} + 2 N_f {C}_{qg}^{'(2)}\Big)
\nn\\
=&\
C_A N_f \left(L_\perp \left(\frac{32 \zeta _2}{3}+\frac{31
   L_Q}{9}-\frac{335}{27}\right)+L_\perp^2 \left(2 L_Q-\frac{44}{9}\right)+\frac{56
   L_Q}{27}\right)
\nn\\
&\ +C_A^2 \left(L_\perp \left(-\frac{176 \zeta _2}{3}+12 \zeta _3+\left(20
   \zeta _2-\frac{205}{9}\right) L_Q+\frac{1069}{27}\right)+L_\perp^2 \left(2 L_Q^2-11
   L_Q+\frac{121}{9}\right)
\right.
\nn\\
&\
\left.
+\left(14 \zeta _3-\frac{404}{27}\right) L_Q\right)-2 C_F
   N_f L_\perp+N_f^2 \left(\frac{4}{9} L_\perp^2+\frac{22}{27} L_\perp\right) + c_2^g \,,
\end{align}
where the two-loop scale-independent constant for gluon jet is
\begin{align}
  \label{eq:cjgluon}
  c_2^g =&\ \left(\frac{109 \zeta _2}{9}+\frac{8 \zeta _3}{3}-\frac{1123}{162}\right) C_A
   N_f+\left(-\frac{751 \zeta _2}{9}-\frac{176 \zeta _3}{3}+135 \zeta
   _4+\frac{2590}{81}\right) C_A^2
\nn\\
&\
+\left(8 \zeta _3-\frac{37}{6}\right) C_F
   N_f-\frac{8 N_f^2}{81} \,.
\end{align}
The gluon jet constant in Eq.~\eqref{eq:cjgluon} enters the back-to-back contact terms in the Higgs EEC at two loops. 
These constants provide important ingredients for N$^3$LL resummation of the EEC in Higgs gluonic decay in the back-to-back limit. They have also been used in Ref.~\cite{Dixon:2019uzg} to extract the collinear contact terms for the Higgs EEC using energy-conservation sum rule. 

We note that the two-loop jet function has also been computed explicitly~\cite{Yangtz} from taking the asymptotic expansion of the differential equation satisfied by the master integrals for the Higgs EEC at finite angle~\cite{Luo:2019nig}. Our results in this paper for the jet function agree perfectly with the brute force calculation in Ref.~\cite{Yangtz}. Another independent check of our results comes from the momentum-conservation sum rule for the EEC, discovered in Refs.~\cite{Kologlu:2019mfz,Korchemsky:2019nzm}. A similar discussion for the quark jet function can be found in Ref.~\cite{Luo:2019hmp}.

We stress that the jet functions are the second moment of the sum of all the TMDFFs, including both the unpolarized and linearly polarized contributions. Therefore, the aforementioned checks also apply to all the TMDFFs presented in this paper. 

It is also interesting to consider the contribution to gluon jet function from linearly polarized gluon alone. They are given by
\begin{align}
  \label{eq:linearpoljet}
  J_1^{g, \rm l.p.} = &\ \frac{C_A}{3}-\frac{N_f}{3} \,,
\nn\\
J_2^{g, \rm l.p.} = &\ 
C_A N_f \left(\frac{8 \zeta _2}{3}+L_\perp \left(\frac{2
   L_Q}{3}-\frac{26}{9}\right)-\frac{163}{27}\right)+C_A^2 \left(-\frac{8 \zeta
   _2}{3}+L_\perp \left(\frac{22}{9}-\frac{2 L_Q}{3}\right)+\frac{107}{27}\right)
\nn\\
&\ +2 C_F
   N_f+N_f^2 \left(\frac{4 L_\perp}{9}+\frac{2}{27}\right) \,.
\end{align}
It's interesting to note that the linearly polarized gluon contribution in Eq.~\eqref{eq:linearpoljet} vanishes in the ${\cal N}=1$ supersymmetry limit, where $C_F = C_A = N_f$. This is simply the consequence of the supersymmetric momentum-conservation sum rule in Eq.~\eqref{eq:FFsumrule}.

Recently, EEC in the back-to-back limit has also been studied at hadron collider, the so-called TEEC. Due to the special geometry at hadron colliders, the linearly polarized gluon does not contribute to TEEC~\cite{Gao:2019ojf}. In that case, the gluon jet function is simply given by 
\begin{align}
  \label{eq:TEECgluonjet}
  J_{\rm TEEC}^g(b_\perp, \mu, \nu, \alpha_s) = 
 J^g(b_\perp, \mu, \nu, \alpha_s) -  J^{g,\rm l.p.}(b_\perp, \mu, \nu, \alpha_s) \,.
\end{align}
Therefore, our results also provide the missing ingredients for the N$^3$LL resummation of the TEEC in the back-to-back limit. 

\section{Conclusion}
\label{sec:conclusion}

In this paper we have presented the perturbative gluon TMD coefficient functions through two loops, for both unpolarized and linearly polarized coefficients. Our calculation was performed using the exponential regulator for the regularization of rapidity divergences. We have obtained results through $\Ord(\e^2)$ at two loops, which are relevant for future three-loop calculation. Our results are rapidity regulator dependent. Rapidity regulator independent results can be obtained by multiplying the TMD soft function properly, see Eq.~\eqref{eq:compareTMDPDF}. We have compared our results with those in the literatures, and found perfect agreement in most cases. However, discrepancy with Ref.~\cite{Gutierrez-Reyes:2019rug} was found for the linearly polarized gluon TMDPDFs at two loops, and with Ref.~\cite{Echevarria:2016scs} for the $\delta(1-z)$ term in the gluon-to-gluon TMDFF. As a by-product of this calculation, we found a momentum conservation sum rule in the ${\cal N}=1$ supersymmetric limit for the linear polarization of gluon TMDPDFs and TMDFFs. Our results provide important ingredients for the precision studies of gluon TMDs in collider experiments.

\acknowledgments

We are grateful to Xing Wang, Xiaofeng Xu, Li Lin Yang for related collaborations. This work was supported in part by the National Natural Science Foundation of China under contract No.~11975200 and the Zhejiang University
Fundamental Research Funds for the Central Universities (2017QNA3007,	107201*172210191).

\appendix

\section{Anomalous dimensions, splitting functions, renormalization factors and the TMD soft function}
\label{sec:appendix}

In this Appendix, we list some necessary ingredients which enter our calculation.

\subsection{Anomalous dimensions}
\label{sec:AD}

For all the anomalous dimensions entering the RGEs of various TMD functions, we define the perturbative expansion according to
\begin{equation}
\label{eq:Aanomalous}
\gamma(\als) = \sum_{n=0}^\infty \left( \alspi \right)^{n+1} \, \gamma_n \,,
\end{equation}
where the coefficients up to $\Ord(\alpha_s^2)$ are
\begin{align}
\Gcusp_{0} &= 4 C_A  \,, \nn
\\
\Gcusp_{1} &= C_A^2 \left(\frac{268}{9}-8 
                 \zeta_2\right)-\frac{80 C_A T_F N_f}{9} \,, \nn
\\
\gamma^R_0 &= 0 \, , \nn
\\
\gamma^R_1 &= C_A \left[ C_A \left( -\frac{404}{27} + 14\zeta_3 \right) + T_FN_f \frac{112}{27} \right] \,, \nn \\
\gamma^B_0 &= \frac{11}{3} C_A - \frac{4}{3} T_F N_f \,, \nn
\\
\gamma^B_1 &= C_A^2 \left( \frac{32}{3}+ 12 \zeta_3 \right) + \left(  -\frac{16}{3} C_A -  4 C_F \right) N_f T_F   \,, \nn
\\
\gamma^H_0 &= - \frac{11}{3} C_A + \frac{4}{3} T_F N_f  \,, \nn
\\
\gamma^H_1 &=  C_A^2 \left(2 \zeta_3+\frac{11 }{3} \zeta_2 -\frac{692}{27}\right) +  C_A  N_f T_F \left(\frac{256}{27}-  \frac{4}{3} \zeta_2  \right)+   4 C_F N_f T_F  \,, \nn
\\
\gamma^S_0 &= 0 \,, \nn
\\
\gamma^S_1 &= C_A \left[ C_A \left( -\frac{404}{27} + \frac{11}{3} \zeta_2 + 14\zeta_3 \right) + T_FN_f \left( \frac{112}{27} - \frac{4 }{3} \zeta_2 \right) \right] \,.
\end{align}
The cusp anomalous dimension $\Gamma^{\text{cusp}}$ can be found in \cite{Korchemsky:1987wg,Korchemskaya:1992je} and rapidity anomalous dimension $\gamma^R$ can be found in Refs.~\cite{Li:2016ctv,Vladimirov:2016dll}. The hard and soft anomalous dimensions $\gamma^H$ and $\gamma^S$ can be extracted from the two-loop gluon form factor~\cite{Gehrmann:2005pd}, and can also be found in, e.g., Refs.~\cite{Becher:2009qa,Li:2014afw}. Finally, the beam anomalous dimension $\gamma^B$ is related to $\gamma^S$ and $\gamma^H$ through $\gamma^B = \gamma^S - \gamma^H$. 

The QCD beta function is defined by
\begin{equation}
\frac{d\als}{d\ln\mu} = \beta(\als) = -2\als \sum_{n=0}^\infty \left( \alspi \right)^{n+1} \, \beta_n \,,
\end{equation}
with \cite{Gross:1973id,Politzer:1973fx,Caswell:1974gg,Jones:1974mm,Egorian:1978zx}
\begin{align}
\beta_0 &= \frac{11}{3} C_A - \frac{4}{3} T_F N_f \,, \nn
\\
\beta_1 &= \frac{34}{3} C_A^2 - \frac{20}{3} C_A T_F N_f - 4 C_F T_F N_f \,,
\end{align}
A formula particularly useful for us is
\begin{equation}
\als(b_0/b_T) = \frac{\alsmu}{t} \left[ 1 - \alsmupi \frac{\beta_1}{\beta_0} \frac{\ln t}{t} \right] + \Ord(\als^3) \, ,
\end{equation}
where
\begin{equation}
t = 1 - \alsmupi \beta_0 \Lp \, .
\end{equation}

\subsection{Space-like splitting functions}

The LO space-like splitting functions can be written as \cite{Altarelli:1977zs}
\begin{align}
P^{\zero}_{qq}(z) &= 2 C_F \left[ p_{qq}(z) +\frac{3}{2} \delta (1-z) \right] \,, \nonumber  \\
P^{\zero}_{gq}(z) &= 2 C_F \, p_{gq}(z) \,, \nonumber \\
P^{\zero}_{gg}(z) &= 4 C_A \, p_{gg}(z) + \delta(1-z) \left( \frac{11}{3}  C_A - \frac{4}{3} T_F N_f \right) \,, \nonumber \\
P^{\zero}_{qg}(z) &=  2 T_F \, p_{qg}(z) \,,
\end{align}
where
\begin{align}
p_{qq}(z) &= \frac{1+z^2}{(1-z)_+} \,, \nonumber \\
p_{gq}(z) &= \frac{1+(1-z)^2}{z} \,, \nonumber \\
p_{gg}(z) &= \frac{z}{(1-z)_+} + \frac{1-z}{z} + z(1-z) \,, \nonumber \\
p_{qg}(z) &= z^2+ (1-z)^2 \,.
\end{align}
The NLO space-like splitting functions are \cite{Furmanski:1980cm,Curci:1980uw}
\begin{align}
P^{\one}_{gq}&=C_A C_F\bigg[-\frac{4}{3}\left(8z^2+15z+36\right) H_{0}+\frac{4}{3}\left( -17z-\frac{22}{z}+22\right) H_{1}-8p_{gq}(-z)H_{-1,0}\nonumber \\
&+ \frac{4}{9z}  \left(44z^3+37z^2+(19+36\zeta_2)z+9 \right) +8p_{gq}(z)\left( H_{1,0}+H_{1,1}+H_{2}\right)+8(z+2)H_{0,0}\bigg]\,,\nonumber \\
P^{\one}_{gg}&=C_A^2\bigg[ \left(\frac{268}{9} - 8 \zeta_2 \right) \frac{1}{(1-z)_+} +8\left(\frac{1}{1-z}-2z^2+4z+\frac{1}{1+z}\right) H_{0,0}\nonumber \\
&+8\bigg(2z^2-\frac{1}{1+z}+4\bigg)\zeta_2-\frac{4}{3}\left( 44z^2-11z+25\right) H_{0}+16p_{gg}(z)\left( H_{1,0}+H_{2}\right)\nonumber \\
&-\frac{2}{9}\left(109z+25\right) + \frac{4}{3}\left( 9\zeta_3+8\right) \delta(1-z) -16p_{gg}(-z)H_{-1,0}\bigg]\nonumber \\
&+N_f C_F T_F\bigg[-16(z+1)H_{0,0}-8(5z+3)H_{0}+\frac{16}{3}\left( 5z^2+6z+\frac{1}{z}-12\right)-4\delta(1-z)
\bigg] \nn \\
&+N_f C_A T_F\bigg[\frac{8}{9z} \bigg( 23z^3-19z^2+29z-23\bigg) - \frac{16}{3} \delta(1-z)-\frac{16(z+1)}{3}H_{0}\bigg]\,.
\end{align}

\subsection{Time-like splitting function}
\label{sec:timelikesplitting}

The LO time-like splitting functions are exactly the same as the space-like ones~(the Gribov-Lipatov reciprocity), while the NLO time-like splitting functions are given by \cite{Furmanski:1980cm, Curci:1980uw}
\begin{align}
P^{T \one}_{qg}&= C_A T_F\bigg[-8p_{qg}(-z)H_{-1,0}+8p_{qg}(z)\left(2H_{1,0}+H_{1,1}-2H_{2}\right)-\frac{8}{3}\left(2z^2+17z+2\right) H_{0}\nonumber \\
&+(48z+8)H_{0,0}-\frac{4}{3}\left(10z^2-10z+11\right)H_{1}-\frac{4}{9z}(-178z^3(95+36\zeta_2)z^2-13z+20)\bigg]\nonumber \\
&+C_F T_F\bigg[ p_{qg}(z)\left(-24H_{1,0}-8H_{1,1}+8H_{2}-8\zeta_2\right)+\left(-16z^2+8z-4\right)H_{0,0}\nonumber \\
&+\left(8z^2+8z-10\right) H_{0}+\left(8z^2-8z+12\right)H_{1}-\frac{2}{3}\left(60z^2-69z+36\right)\bigg] \nonumber \\
&+N_f T_F^2\bigg[ \frac{16}{3}\left(H_{1}-H_{0}\right) p_{qg}(z)-\frac{16}{9}\left(4z^2-4z+5\right)\bigg]\,,\nonumber \\
P^{T \one}_{gg}&=C_A^2\bigg[\left(\frac{268}{9} - 8 \zeta_2 \right) \frac{1}{(1-z)_+} +8\left(-\frac{3}{1-z} +2z^2-8z+\frac{1}{1+z}-\frac{4}{z}\right) H_{0,0}\nonumber \\
&+\frac{4}{3}\bigg(\frac{22}{1-z}-22z^2-3z-\frac{22}{z}-33\bigg)H_{0}+8\bigg(2z^2-\frac{1}{1+z}+4\bigg)\zeta_2\nonumber \\
&-\frac{2}{9}\left(109z+25\right)-16p_{gg}(-z)H_{-1,0}+\frac{4}{3}(9\zeta_3+8)\delta(1-z)-16p_{gg}(z)\left( H_{1,0}+H_{2}\right)\bigg]\nonumber \\
&+N_f C_AT_F\bigg[\frac{80}{9} \frac{1}{(1-z)_+}+ \frac{16(2z^3-3z^2+3z-2)}{3z}H_{0}-\frac{16}{3}\delta(1-z)\nonumber \\
&+\frac{8(23z^3-19z^2+29z-23)}{9z} - \frac{32}{3(1-z)} H_0 \bigg]\nonumber \\
&+N_fC_FT_F\bigg[16(z+1)H_{0,0}+\frac{8}{3}\bigg(8z^2+21z+\frac{8}{z}+15\bigg)H_{0}-4\delta(1-z)\nonumber \\
&+\frac{16}{9}\bigg(-41z^2+27z+\frac{23}{z}-9\bigg)\bigg] \,.
\end{align}

\subsection{Renormalization factors}
Similar with Eq.~\eqref{eq:Aanomalous}, we define all the renormalization factors with the perturbative expansion
\begin{align}
\label{eq:Sreno}
Z(\alpha_s) = 1+ \sum^{\infty}_{n=1} \left( \frac{\alpha_s}{4 \pi}\right)^{n} Z_n \,.  
\end{align}
In addition to the identical renormalization factors $Z_g^B(b_\perp, \mu, \nu)$ for TMDPDFs and TMDFFs, the soft renormalization factor $Z_g^S(b_\perp,\mu,\nu)$ also enter our calculation,
\begin{align}
\mathcal{S}_{gg}^{\text{bare}}(b_\perp, \nu) = Z_g^S(b_\perp, \mu, \nu) \mathcal{S}_{gg}(b_\perp,\mu,\nu) \,.
\end{align}
The first two orders of $Z_g^B $ and $Z_g^S$ are 
\begin{align}
\label{eqZqZs}
Z^B_1 &= \frac{1}{2\epsilon} \left(2 \gamma^B_0 - \Gamma_0^{\text{cusp}} L_Q \right) \,, \nonumber \\
Z^B_2 &= \frac{1}{8 \epsilon^2} \bigg( ( \Gamma_0^{\text{cusp}} L_Q - 2 \gamma^B_0)^2 + 2 \beta_0 (  \Gamma_0^{\text{cusp}} L_Q - 2 \gamma^B_0)    \bigg) + \frac{1}{4\epsilon} \left( 2 \gamma^B_1 - \Gamma_1^{\text{cusp}} L_Q \right) \,, \nonumber \\
Z^S_1 &= \frac{1}{\epsilon^2} \Gamma^{\text{cusp}}_0  +  \frac{1}{\epsilon} \left( -2 \gamma^S_0 - \Gamma_0^{\text{cusp}} L_\nu \right) \,,\nonumber \\
Z^S_2 &= \frac{1}{2 \epsilon^4} (\Gamma^{\text{cusp}}_0)^2 - \frac{1}{4 \epsilon^3} \bigg(\Gamma^{\text{cusp}}_0 (3 \beta_0 + 8 \gamma^S_0)+4( \Gamma^{\text{cusp}}_0)^2 L_\nu\bigg)   - \frac{1}{2 \epsilon} \left( 2 \gamma^S_1 +  \Gamma^{\text{cusp}}_1 L_\nu \right) \nonumber \\
& + \frac{1}{4 \epsilon^2} \bigg(\Gamma^{\text{cusp}}_1 + 2 ( 2 \gamma^S_0 + \Gamma^{\text{cusp}}_0 L_\nu ) ( \beta_0 + 2 \gamma^S_0 + \Gamma^{\text{cusp}}_0 L_\nu) \bigg) \,.
\end{align}

\subsection{Renormalized TMD soft function}
\label{sec:reTMDsoft}
The exponentially regularized TMD soft function is given by \cite{Li:2016ctv}
\begin{align}
\label{eq:soft}
{\cal S}_{gg}(b_\perp,\mu,\nu) =  \exp \bigg\{ \alspi \left( c_1^s+ c_1^{\perp} \right)  +  \left(\alspi\right)^2 \left( c_2^s + c_2^\perp \right) \bigg\}  \,,
\end{align}
where the scale-dependent terms are 
\begin{align}
c_1^s &= \frac{\Gcusp_0}{2} \Lp^2 - \Lp \big( \Gcusp_0 L_R +2\gamma_0^S \big) + 2 \gamma_0^R L_R \,, \nonumber \\
c_2^s &= \frac{\beta_0\Gcusp_0}{6} \Lp^3 + \left( \frac{\Gcusp_1}{2} - \frac{\beta_0\Gcusp_0}{2} L_R - \beta_0 \gamma_0^S \right) \Lp^2 \nonumber 
\\
&+ \left( \left( 2\beta_0 \gamma_0^R - \Gcusp_1 \right) L_R - 2\gamma_1^S + \beta_0 c_1^{\perp} \right) \Lp + 2\gamma_1^R L_R \,, 
\end{align}
where $L_R = \Lp + L_{\nu}$ with $L_\nu = \ln(\nu^2/\mu^2)$. The scale-independent terms are
\begin{align}
c_1^{\perp} &= -2 C_A \zeta_2 \,, \nonumber \\
c_2^{\perp} &= C_A^2 \left( -\frac{67}{3} \zeta_2 - \frac{154}{9} \zeta_3 + 10 \zeta_4 + \frac{2428}{81} \right) + C_A N_f \left( \frac{10}{3} \zeta_2 + \frac{28}{9} \zeta_3 -\frac{328}{81} \right) \,.
\end{align}

\subsection{Bare TMD soft function}
\label{sec:zerobinsoft}
The bare zero-bin soft function is equivalent to bare TMD soft function, 
\begin{align}
\mathcal{S}_{0 \rm b}(b_\perp , \nu) = \mathcal{S}^{\text{bare}}_{gg}(b_\perp, \nu) = Z_g^{S}(b_\perp, \mu,\nu) \left[ \mathcal{S}_{gg}(b_\perp , \mu, \nu)|{c_1^s \to c_1^{s'} , c_1^\perp  \to c_1^{\perp '}} \right] \,,
\end{align}
where we need to expand $c_1^s$ and $c_1^\perp $ to order $\mathcal{O}(\epsilon^2)$,
\begin{align}
c_1^{s'} &= c_1^s +\epsilon C_A  \left(\frac{2
   L_\perp^3}{3}-2 L_R L_\perp^2 -2 \zeta _2 \left( L_\perp + L_R\right)\right) \nonumber \\
   &+ \epsilon ^2 C_A \left(\frac{L_\perp^4}{6} -\frac{2}{3} L_R L_\perp^3
   - \zeta _2 \left(2 L_\perp L_R+L_\perp^2\right)- \frac{4}{3} \zeta _3 \left(2 L_\perp+L_R\right)\right) \,,\nonumber \\
   c_1^{\perp '} &= c_1^{\perp} -  C_A   \left( \epsilon \frac{8}{3} \zeta_3 + \epsilon^2  \frac{27}{4}\zeta_4  \right) \,.   
\end{align}

\bibliographystyle{JHEP}
\bibliography{twoloop_gluon}

\end{document}